\documentclass[%
superscriptaddress,
longbibliography,
preprint,
 amsmath,amssymb,
 aps,
floatfix,
]{revtex4-1}

\usepackage{graphicx}% Include figure files
\usepackage{dcolumn}% Align table columns on decimal point
\usepackage{bm}% bold math
\usepackage{color,amssymb,colortbl,subfigure,graphicx} 
\usepackage[table]{xcolor}
\newcommand{\tm}{t_{\rm{mod}}}
\newcommand{\thd}{t_{\rm{hold}}}

\usepackage{lineno}
\begin{document}

%\preprint{APS/123-QED}
%\preprint{APS/123-QED}

\title{Parametric Excitation of a Bose-Einstein Condensate:  From Faraday Waves to Granulation}% Force line breaks with \\

\author{J. H. V. Nguyen}
\affiliation{Department of Physics and Astronomy, Rice University, Houston, Texas 77005, USA}
\author{M. C. Tsatsos}
\affiliation{Institute of Physics of S\~ao Carlos, University of S\~ao Paulo, PO Box 369, 13560-970, S\~ao Carlos, SP, Brazil}
\author{D. Luo}
\affiliation{Department of Physics and Astronomy, Rice University, Houston, Texas 77005, USA}
\author{A. U. J. Lode}
\affiliation{Wolfgang Pauli Institute c/o Faculty of Mathematics, University of Vienna, Oskar-Morgenstern Platz 1, 1090 Vienna, Austria}
\affiliation{Vienna Center for Quantum Science and Technology, Atominstitut, TU Wien, Stadionallee 2, 1020 Vienna, Austria}
\affiliation{Department of Physics, University of Basel, Klingelbergstrasse 82, CH-4056 Basel, Switzerland}
\author{G. D. Telles}
\affiliation{Institute of Physics of S\~ao Carlos, University of S\~ao Paulo, PO Box 369, 13560-970, S\~ao Carlos, SP, Brazil}
\author{V. S. Bagnato}
\affiliation{Institute of Physics of S\~ao Carlos, University of S\~ao Paulo, PO Box 369, 13560-970, S\~ao Carlos, SP, Brazil}
\author{R. G. Hulet}
\affiliation{Department of Physics and Astronomy, Rice University, Houston, Texas 77005, USA}

\date{\today}% It is always \today, today,

\begin{abstract}
We explore, both experimentally and theoretically, the response of an elongated Bose-Einstein condensate to modulated interactions.  We identify two distinct regimes differing in modulation frequency and modulation strength.   Longitudinal surface waves are generated either resonantly or parametrically for modulation frequencies near the radial trap frequency or twice the trap frequency, respectively.  The dispersion of these waves, the latter being a Faraday wave, is well-reproduced by a mean-field theory that accounts for the 3D nature of the elongated condensate.  In contrast, in the regime of lower modulation frequencies we find that no clear resonances occur, but with increased modulation strength, the condensate forms an irregular granulated distribution that is outside the scope of a mean-field approach.  We find that the granulated condensate is characterized by large quantum fluctuations and correlations, which are well-described with single-shot simulations obtained from wavefunctions computed by a beyond mean-field theory at zero temperature, the multiconfigurational time-dependent Hartree for bosons method.

%\begin{description}
%\end{description}
\end{abstract}

\pacs{Valid PACS appear here}% PACS, the Physics and Astronomy
\maketitle

%\tableofcontents

\section{\label{sec:intro} Introduction}

Spatial patterns frequently emerge in driven fluids in a variety of contexts, including chemistry, biology, and nonlinear optics~\cite{Cross1993}.  Instabilities in these systems can generally be categorized as Rayleigh-B\'{e}nard convection, Taylor-Couette flow, or parametric surface waves.  One of the earliest and best known examples of the latter type are the surface waves found by Faraday when a vessel containing a fluid was shaken vertically~\cite{Faraday1831}.  The resulting standing wave patterns that appear on the fluid surface arise from parametric excitation of collective modes of the fluid.  The Faraday experiment has been repeated in various geometries where complex patterns were observed for small driving amplitudes~\cite{Douady1988}.  Chaotic behavior, such as sub-harmonic bifurcation, is seen when the drive amplitude is strong~\cite{Keolian1981,Ciliberto1984,Douady1988,Ciliberto1991} and this behavior has been connected to the onset of turbulence~\cite{Feigenbaum1979}.

A model of the Faraday instability has been developed for an inviscid fluid in which the underlying hydrodynamic equations have been linearized~\cite{Benjamin1954}.  The linearized dynamics are described by a Mathieu equation, $\ddot{x} + p(t)x = 0$, where $x$ is the displacement, $p(t) = \Omega^2 (1 + \epsilon \cos(\omega t))$ is the drive, representing a parametrically driven (undamped) harmonic oscillator with a natural frequency $\Omega$, drive frequency $\omega$, and drive amplitude $\epsilon$.  Solving the equations using a Floquet analysis results in a series of  resonances at $\omega = 2\Omega / n$, where $n$ is an integer~\cite{Bechoefer1996}.

Superfluids are particularly interesting in the context of Faraday waves because the damping of collective modes can be much smaller than in normal fluids, and because patterns may dissipate by the formation of quantized vortices in two or three dimensions.  Several theoretical works have investigated Faraday waves in Bose-Einstein condensates (BECs) of atomic gases~\cite{Garcia-Ripoll1999, Staliunas2002, Staliunas2004, Nicolin2007, Nath2010, Nicolin2011, Balaz2014}.  To our knowledge, only three experiments on Faraday waves in superfluids have been performed, one in which a vessel containing liquid $^4$He is vertically shaken in a way similar to the original Faraday experiment~\cite{Abe2007}, a pioneering experiment in which Faraday waves were excited by modulation of the transverse trap frequency, $\omega_r$, of an elongated BEC of Rb atoms~\cite{Engels2007}, and another in which a non-destructive imaging technique was used to observe Faraday waves in a BEC of Na atoms~\cite{Groot2015}.  In the BEC experiments, the transverse breathing mode, excited at a frequency of $2\omega_r$, strongly couples to the density, and hence, to the nonlinear interactions of the condensate.  This coupling produces the longitudinal sound waves responsible for creating Faraday waves~\cite{Engels2007,Groot2015}.  The spatial period of the Faraday waves was measured as a function of $\omega$, and the response to the strength $\epsilon$ of the drive was investigated~\cite{Engels2007}.  In a related BEC experiment, modulation of the scattering length in a regime of large modulation amplitude and frequency resulted in the stimulated emission of matter-wave jets from a $2$D BEC of Cs atoms~\cite{Clark2017}.  

In this paper, we report measurements characterizing the response of an elongated BEC to direct modulation of the interaction parameter using a Feshbach resonance~\cite{Malomed2006, Pollack2010, Vidanovic2011}.  For  drive frequencies near the first parametric resonance ($\omega$ near $2\omega_r$), we observe robust linear spatial patterns characterized by a spatial period $\lambda_F$($\omega$) consistent with Faraday waves.  We also observe the response of the gas to the next lowest ``resonant'' mode ($\omega$ near $\omega_r$)~\cite{Nicolin2011}.  We have also investigated how $\lambda_F$ depends on the interaction strength.  These measurements are compared with a theory that fully incorporates radial, as well as axial dynamics using a variational method~\cite{Nicolin2011}, and, as we will show, the agreement is excellent.

We also explore a different modulation regime, both experimentally and theoretically, where $\omega$ is far from any trap frequency.  The behavior in this regime is distinctly different; no clear resonances are observed, and much larger $\epsilon$ and modulation times are needed to obtain a significant response.  The response is not regular in this regime, and no clear patterns emerge; rather, modulation produces a series of irregular grains.  

Granulation is found in a variety of systems extending over many length and energy scales~\cite{Jaeger1996, Mehta1994}. In quantum gases, granular states have been discussed previously in the context of perturbed atomic BECs and explored theoretically using a mean-field approach~\cite{Yukalov2014, Yukalov2015}.  Granular states have been defined to have the following properties~\cite{Yukalov2014}:  i) they are dynamical quantum states where particles cluster in higher density grains interleaved by regions of very low density, ii) the spatial distribution of grains is random, and iii) the grain size is variable and of a multiscale nature.

Our theoretical description uses the multiconfigurational time-dependent Hartree method for bosons (MCTDHB)~\cite{Streltsov2007,Alon2008}.  MCTDHB captures many of the salient experimental observations and goes systematically beyond a mean-field description obtained from the Gross-Pitaevskii equation.  The discrepancies between the Gross-Pitaevskii mean-field description and both the experimental observations, and our MCTDHB results hint that granulation emerges concurrently with many-body correlations.

\section{Faraday waves}
In our experiment, we confine a gas of up to $8 \times 10^5$ $^7$Li atoms in a single-beam optical dipole trap and cool them to well-below $T_c$, the transition temperature for Bose-Einstein condensation~\cite{Pollack2010}.  This configuration results in a highly elongated cylindrical trapping geometry whose corresponding axial and radial harmonic frequencies are $\omega_z = (2\pi) 7\ \mathrm{Hz}$ and $\omega_r = (2\pi) 475\ \mathrm{Hz}$, respectively.  The atoms are optically pumped into the lowest ground state hyperfine level, ${|F = 1, m_F = 1\rangle}$, where their $s$-wave scattering length may be controlled using a broad Feshbach resonance located at $737.7\ \mathrm{G}$~\cite{Pollack2009, Gross2011, Navon2011, Dyke2013}.  The magnetic field is sinusoidally modulated according to $B(t) = \bar{B} + \Delta B \sin(\omega t)$, resulting in a modulated scattering length, $a(t)$.  The modulation amplitude $\Delta B$, modulation time $t_m$ and hold time $t_h$ following $t_m$ are varied for each value of the modulation frequency $\omega$, as necessary to produce a Faraday pattern with similar contrast.  After $t_h$, we take a polarization phase contrast image~\cite{Bradley1997} with a probe laser propagating along the $x$-axis, perpendicular to the cylindrical $z$-axis of the trap.  These images provide column density distributions that we integrate along the $y$-axis to obtain line density profiles.  We apply a fast-Fourier transform (FFT) to these profiles in order to determine the spectrum of spatial frequencies exhibited by the BEC following modulation.

\begin{figure}
\includegraphics[width=\linewidth]{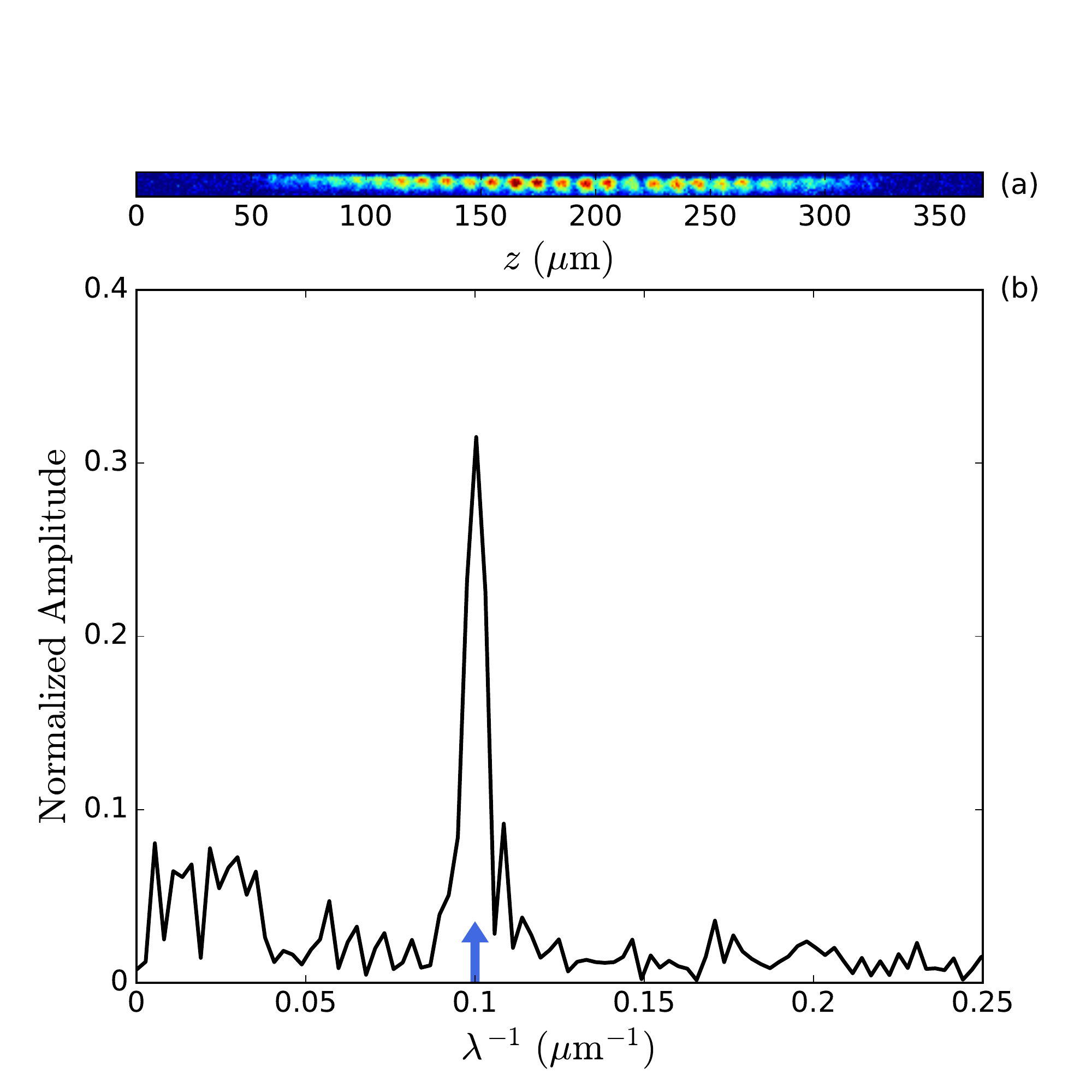}
%\captionsetup[figure]{format=plain, justification=justified}
\caption{(a): column density image; (b):  FFT of the line density.  The modulation parameters are:  $\omega = (2\pi)950$ Hz, $\bar{B} = 572.5$ G, $\Delta B = 5$ G, corresponding to a mean scattering length $\bar{a} = 4.2 a_0$, and a modulation amplitude $\Delta a = 0.9 a_0$, where $a_0$ is the Bohr radius.  In addition, $t_m = 5\ \mathrm{ms}$ and $t_h = 20\ \mathrm{ms}$.  The blue arrow indicates the calculated $\lambda_F^{-1}$ for these parameters.  The DC component has been subtracted, and the FFT amplitude is normalized by this DC value.}
\label{fig:fig1}
\end{figure}

A typical image of a single experimental run is shown in Fig.~\ref{fig:fig1}(a).  In this example, $\omega=(2\pi)950\ \mathrm{Hz}$ is resonant with the Faraday mode at $\omega=2\omega_r$.  A surface wave is generated after $t_m = 5\  \mathrm{ms}$ of modulation followed by $t_h = 20\  \mathrm{ms}$.  The FFT, shown in Fig.~\ref{fig:fig1}(b), features a single dominant peak corresponding to a spatial period of $\lambda = 10\ \mathrm{\mu m}$.   

\begin{figure}
\includegraphics[width=\linewidth]{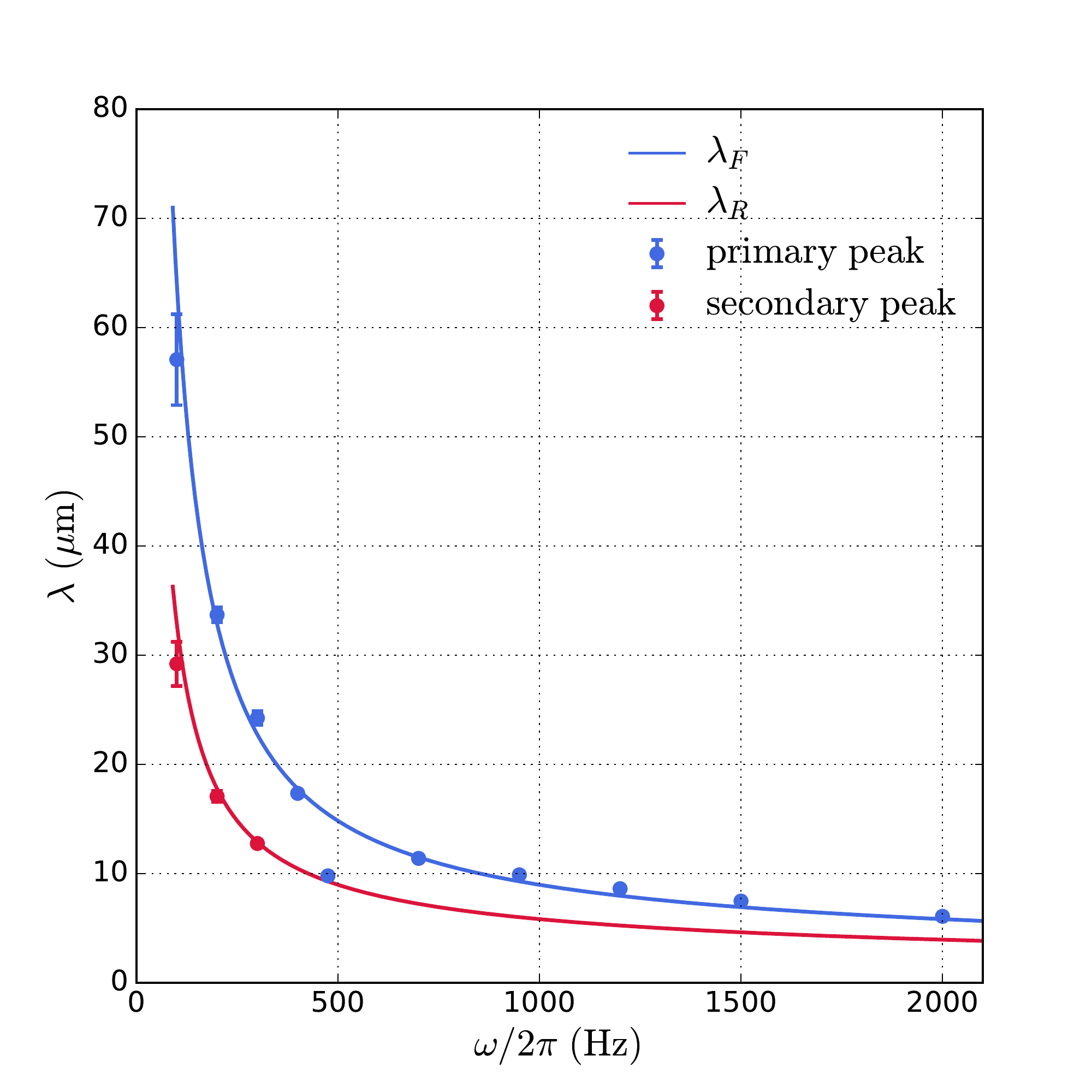}
\caption{Spatial period vs.~$\omega$.  The blue data points are the primary peak of the FFT’s, while the red data points correspond to a secondary peak, where one exists.  The error bars here, and in each subsequent figure, corresponds to the standard error of the mean determined from $10$ different experimental runs for each point.  The solid blue line is the calculated spatial period $\lambda_F$ of the Faraday mode, while the red is that of the resonant mode $\lambda_R$~\cite{Nicolin2011}.  The resonant mode only dominates when $\omega$ is tuned to resonance at $\omega_r$, producing the observed primary peak.}
\label{fig:fig2}
\end{figure}

Figure~\ref{fig:fig2} shows the spatial period of the observed structure as a function of $\omega$.  Typically, $t_h=0\ \mathrm{ms}$ and $20<t_m<40\ \mathrm{ms}$, with the exception of $\omega = \omega_r$ and $\omega = 2\omega_r$.  Near these resonances, the modulation time was kept short, $t_m = 20\ \mathrm{ms}$ and $t_m=5\ \mathrm{ms}$, respectively, followed by $t_h=20\ \mathrm{ms}$.  The blue data points correspond to the spatial period of the primary peak in the FFT spectrum.  Except for the point at $\omega =(2\pi)475\ \mathrm{Hz}$, the period monotonically increases with decreasing $\omega$.  The blue line in Fig.~\ref{fig:fig2} is the result of a $3$D variational calculation of $\lambda_F$~\cite{Nicolin2011}, which fits the data well.  We have verified that the standing wave surface wave amplitude oscillates at $\omega /2$ for $\omega$ near $2\omega_r$, consistent with its identification as a Faraday wave, which is excited parametrically.  

The excitation at $\omega = (2\pi)475$ Hz = $\omega_r$ is not a sub-harmonic of the Faraday mode at $2\omega_r$, but rather the next lowest mode in the infinite series of modes, identified as the ``resonant'' mode in Ref.~\citenum{Nicolin2011}.  In addition to having a different dispersion relation, this mode is also weaker, and therefore more difficult to excite, except exactly on resonance, $\omega = \omega_r$, where the growth rate of the resonant mode exceeds that of the Faraday mode~\cite{Nicolin2011}.  A similar excitation at $\omega_r$ was previously reported \cite{Engels2007}.  The theoretical calculation of the period of this mode is indicated in Fig.~\ref{fig:fig2} by the red line, $\lambda_R$~\cite{Nicolin2011}.  

\begin{figure}
\includegraphics[width=\linewidth]{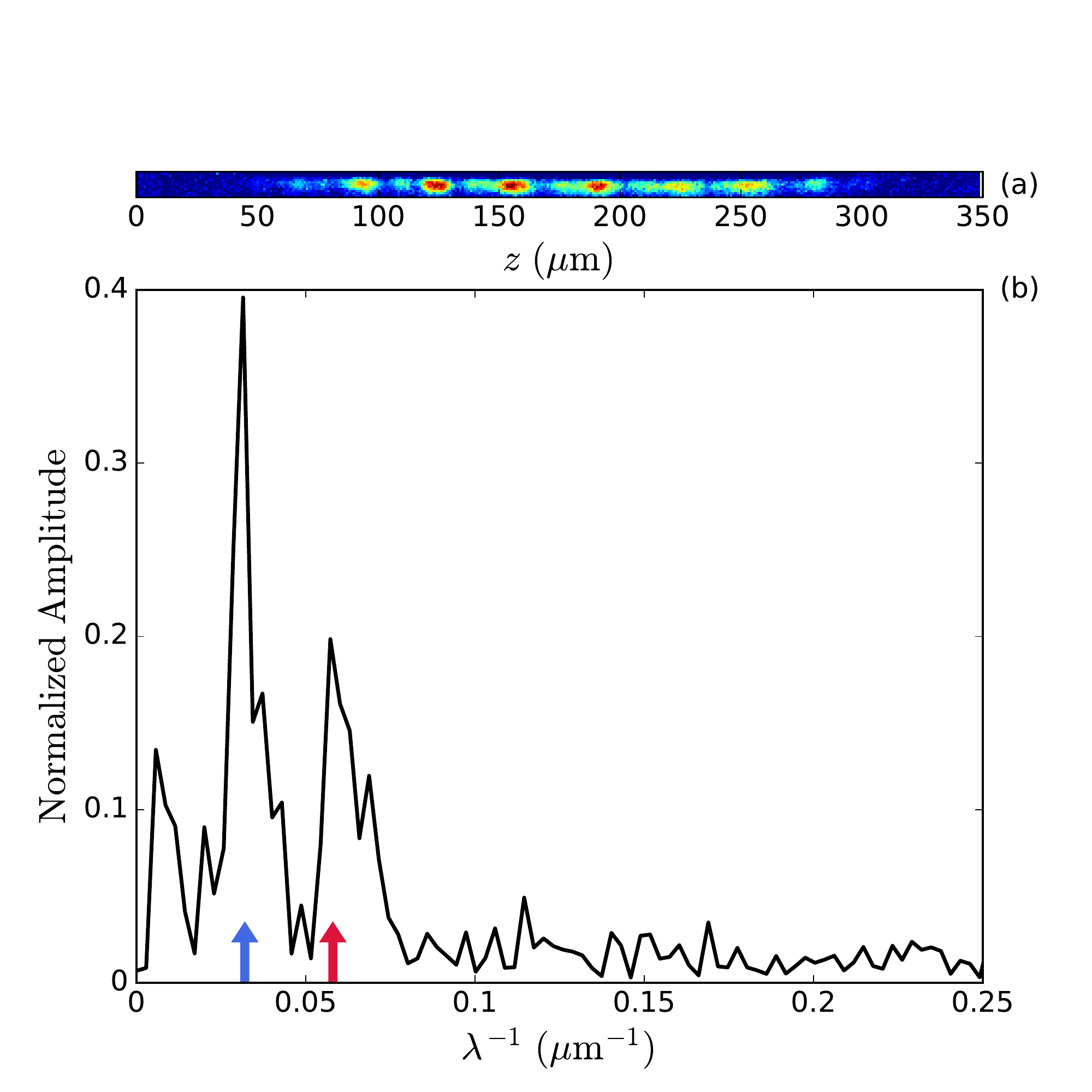}
\caption{(a): Image at $\omega = (2\pi)200\ \mathrm{Hz}$.  (b):  Spectrum showing the primary peak, which corresponds to $\lambda_F$, and the secondary peak due to the resonant mode. The blue and red arrows indicate the calculated values for $\lambda_F^{-1}$ and $\lambda_R^{-1}$, respectively, for these parameters.  Here, $\Delta B = 35\ \mathrm{G}$, but since $a(B)$ is a nonlinear function of $\Delta B$, the bounds $a_{+}=12a_0$ and $a_{-}=-0.9a_0$ are not symmetrically located about $\bar{a}=4.2a_0$.  Also, $t_m=t_h=20\ \mathrm{ms}$.}
\label{fig:fig3}
\end{figure}

We find that as $\omega$ is tuned away from $2 \omega_r=(2\pi)950\ \mathrm{Hz}$, a larger modulation amplitude $\Delta B$ and modulation time $t_m$ are required to obtain a pattern with similar contrast.  For example, Fig.~\ref{fig:fig3} displays the spectrum for $\omega = (2\pi)200\ \mathrm{Hz}$, for which $\Delta B = 35\ \mathrm{G}$, $t_{m} = 20\ \mathrm{ms}$, and $t_h=20\ \mathrm{ms}$.  Two peaks dominate the spectrum: the primary peak at lower spatial frequency, and a secondary peak at roughly twice this spatial frequency.  These secondary peaks only appear for $\omega \lesssim (2\pi) 400\ \mathrm{Hz}$, and are identified by the red data points in Fig.~\ref{fig:fig2}.  The appearance of the next lowest mode depends on being sufficiently near its resonance frequency at $\omega = \omega_r$, and far enough off-resonant with the Faraday mode at $\omega=2\omega_r$ that it does not dominate the FFT spectrum.  We have looked for additional modes in the data, but the FFT spectrum is dominated by the off-resonant response to the $2\omega_r$ and $\omega_r$ resonances, and we are unable to observe any resonances below $\omega_r$.  A comparison of the period of these secondary peaks with the theoretically calculated solid red line indicates that they correspond to the resonant mode $\lambda_R$.

We also explored a more impulsive regime, with short $t_m$, and where $\omega$ is kept within $10\%$ of the Faraday resonance at $\omega=2\omega_r$.  In this case, with short $t_m$, we find that the wavelength of the resulting Faraday pattern is constant, independent of $\omega$.

\begin{figure}
\includegraphics[width=\linewidth]{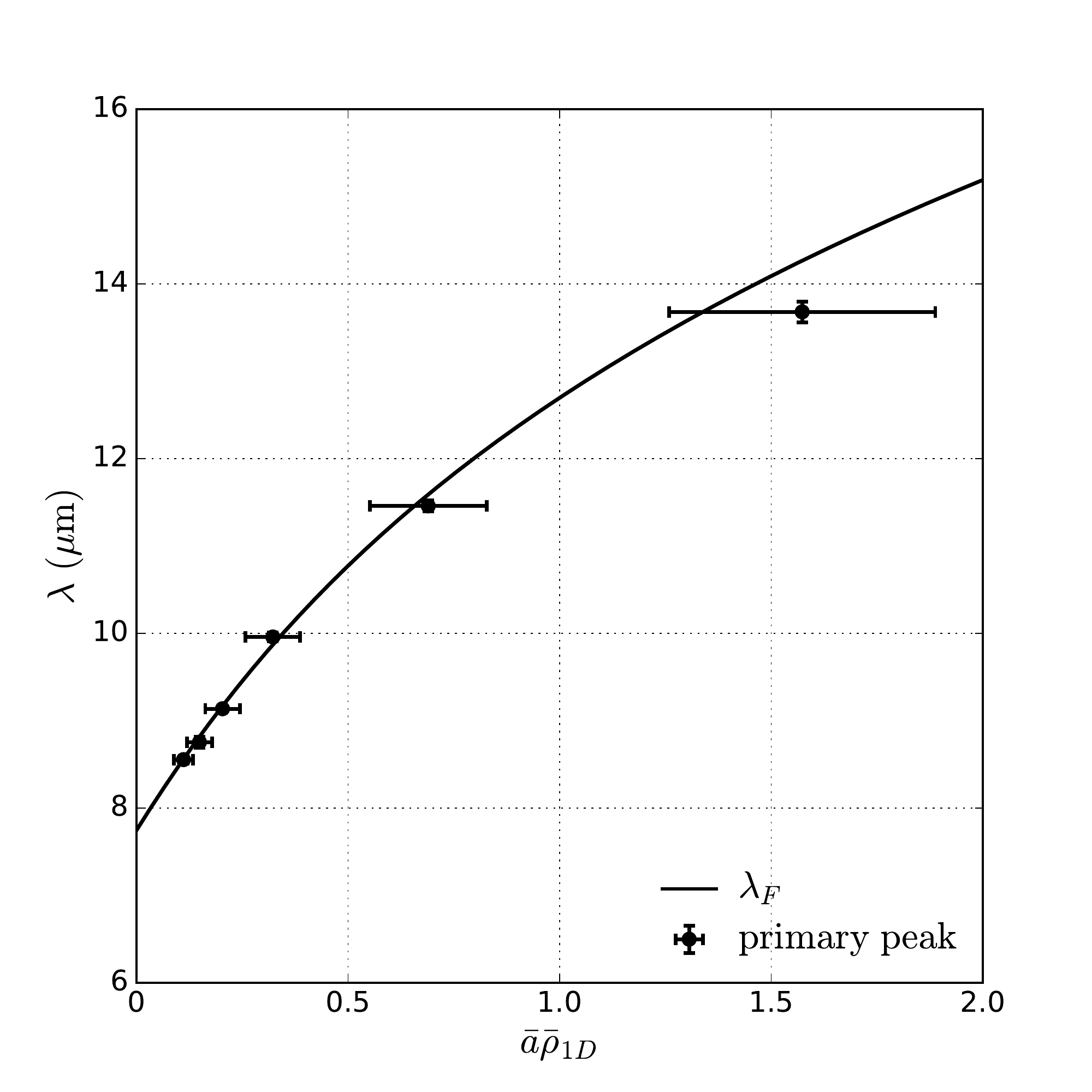}
\caption{Interaction dependence of $\lambda_F$.  The relevant interaction parameter is $\bar{a}\bar{\rho}$, where $\bar{\rho}$ is the average line density and $\bar{a}$ varies between $1a_0$ and $26a_0$.  Here, $\Delta B= 5\ \mathrm{G}$, corresponding to $\Delta a = 0.7a_0$ for $\bar{a}=1a_0$ and $\Delta a = 3a_0$ for $\bar{a}=26a_0$.   The data are indicated by filled squares, while the solid line is the theory of Ref.~\cite{Nicolin2011}.  The error bars along the vertical axis correspond to the standard error, determined from $10$ different experimental runs while the error bars along the horizontal axis arise from the systematic uncertainty in determining $\bar{a}$~\cite{Pollack2009}.}
\label{fig:fig4}
\end{figure}

The Faraday period also depends on the strength of the nonlinearity, as shown in Fig.~\ref{fig:fig4}, where both the measured and calculated~\cite{Nicolin2011} values of $\lambda_F$ are plotted vs.~the interaction parameter $\bar{a}\bar{\rho}$, where $\bar{\rho}$ is the line density obtained by integrating the column density along the transverse direction.  The measured period is consistent with the $3$D theory from Ref~\cite{Nicolin2011}.

\begin{figure}
\includegraphics[width=\linewidth]{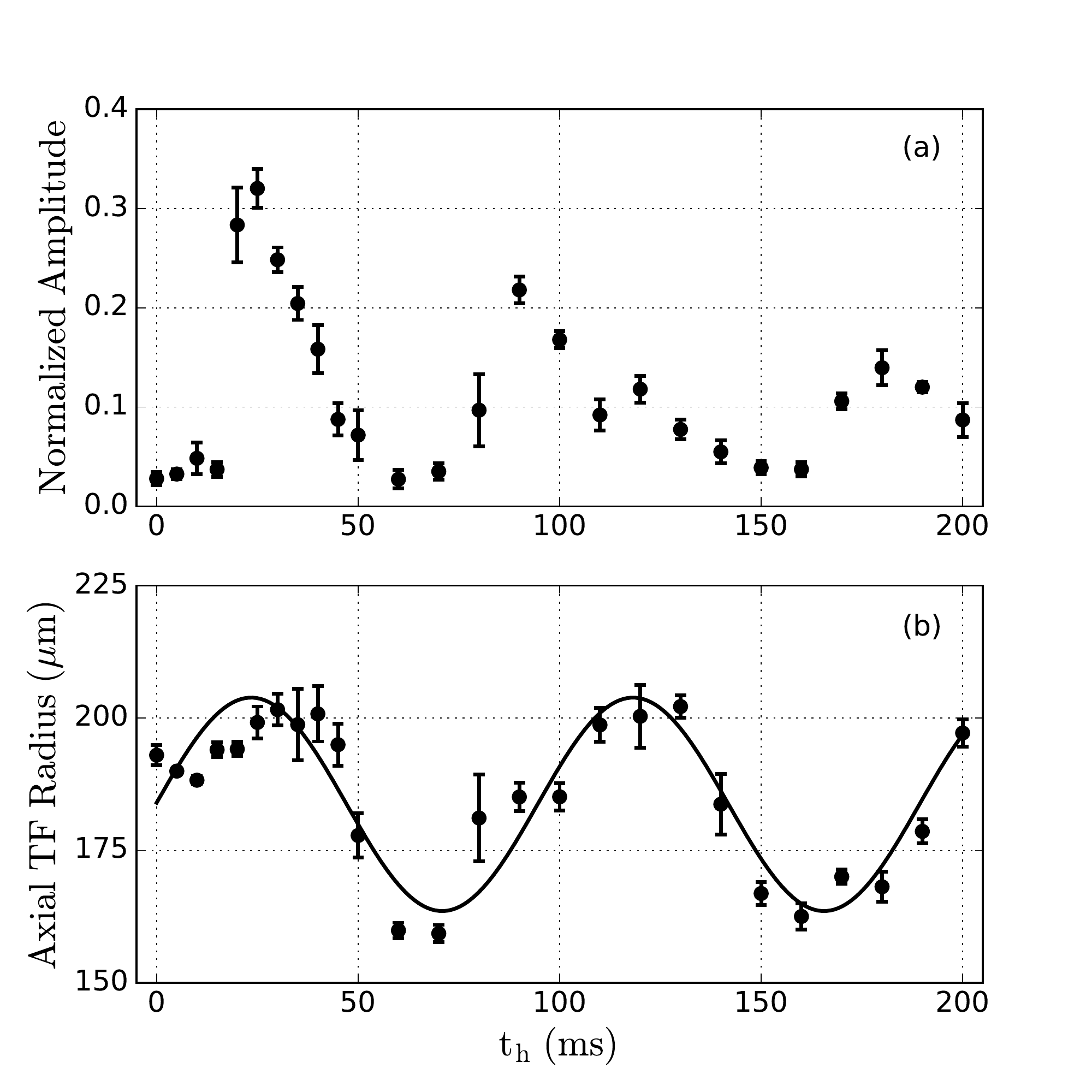}
\caption{Growth and suppression of the Faraday pattern. (a) The normalized amplitude of the primary spatial frequency in the FFT spectrum as function of $t_h$. (b) The fitted axial Thomas-Fermi radius of the central region over the same time interval are shown by the filled circles.  The solid line is a sinusoidal fit corresponding to a period of $95\ \mathrm{ms}$.  For this data, $\omega = (2\pi)950\ \mathrm{Hz}$ and $t_m = 5\ \mathrm{ms}$.}
\label{fig:fig5}
\end{figure}

We have also explored the dynamics for the emergence of the Faraday pattern and its persistence following a short modulation time interval of $t_m = 5\ \mathrm{ms}$ near $2\omega_r$. Figure \ref{fig:fig5}(a) shows the magnitude of the primary peak in the FFT spectrum vs.~$t_h$.  Following modulation, the Faraday pattern forms after $t_h \simeq 20\ \mathrm{ms}$.  By $t_h = 50\ \mathrm{ms}$, however, the Faraday pattern vanishes before reemerging again at $t_h \simeq 90\ \mathrm{ms}$.  A subsequent weaker collapse and revival occur at later $t_h$.  We can gain some intuition as to the origins of this behavior by comparing measurements of the condensate length vs.~$t_h$.  Figure~\ref{fig:fig5}(b) shows the axial Thomas-Fermi radius during the same $t_h$ interval.  It shows that a low frequency collective mode is excited by the coupling to the modulated nonlinearity.  The parameters of this condensate place it between the $1$D mean-field and the $3$D cigar regimes~\cite{Menotti2002}.  In the $3D$ Thomas-Fermi limit, the lowest $m=0$ quadrupolar mode for an elongated condensate has a frequency of $\sqrt{5/2}\ \omega_z$ while in the $1$D limit the collective mode oscillates at $\sqrt{3}\omega_z $~\cite{Stringari1996, Mewes1996,Menotti2002}.  For $\omega_z = (2\pi)7\ \mathrm{Hz}$, the corresponding period for this mode is, therefore, ${\sim}90\ \mathrm{ms}$, which is close to the observed oscillation period of $95\ \mathrm{ms}$.  We find that the Faraday pattern is suppressed during axial compression, but subsequently revives as the condensate returns to its original size.  The phase of the two oscillations, the FFT amplitude and the Thomas-Fermi radius, do not exactly coincide.  We attribute this observation to the delay in the initial growth of the Faraday pattern.  We have determined experimentally that the frequency of the collapse and revival of the Faraday pattern scales with the axial trap frequency.  A similar collapse and revival of the Faraday wave was previously observed~\cite{Groot2015}.

\section{Granulation}

A Faraday pattern is not observed for low frequency modulation, for which $\omega \ll \omega_r$.  We find that as $\omega$ is reduced both modulation time $t_m$ and modulation amplitude $\Delta B$ must be increased in order to observe any change.  As these parameters are increased, more spatial frequencies contribute (see Fig.~\ref{fig:fig3}), and as $t_m$ and $\Delta B$ are increased further, we observe random patterns spanning a broad spatial frequency range, resembling grains~\cite{Yukalov2014, Yukalov2015}.  We do not observe a significant thermal fraction before, nor after modulation, and therefore we attribute the observed granular patterns to quantum fluctuations and use a theory applicable to pure states.

\begin{figure}
\includegraphics[width=0.49\textwidth]{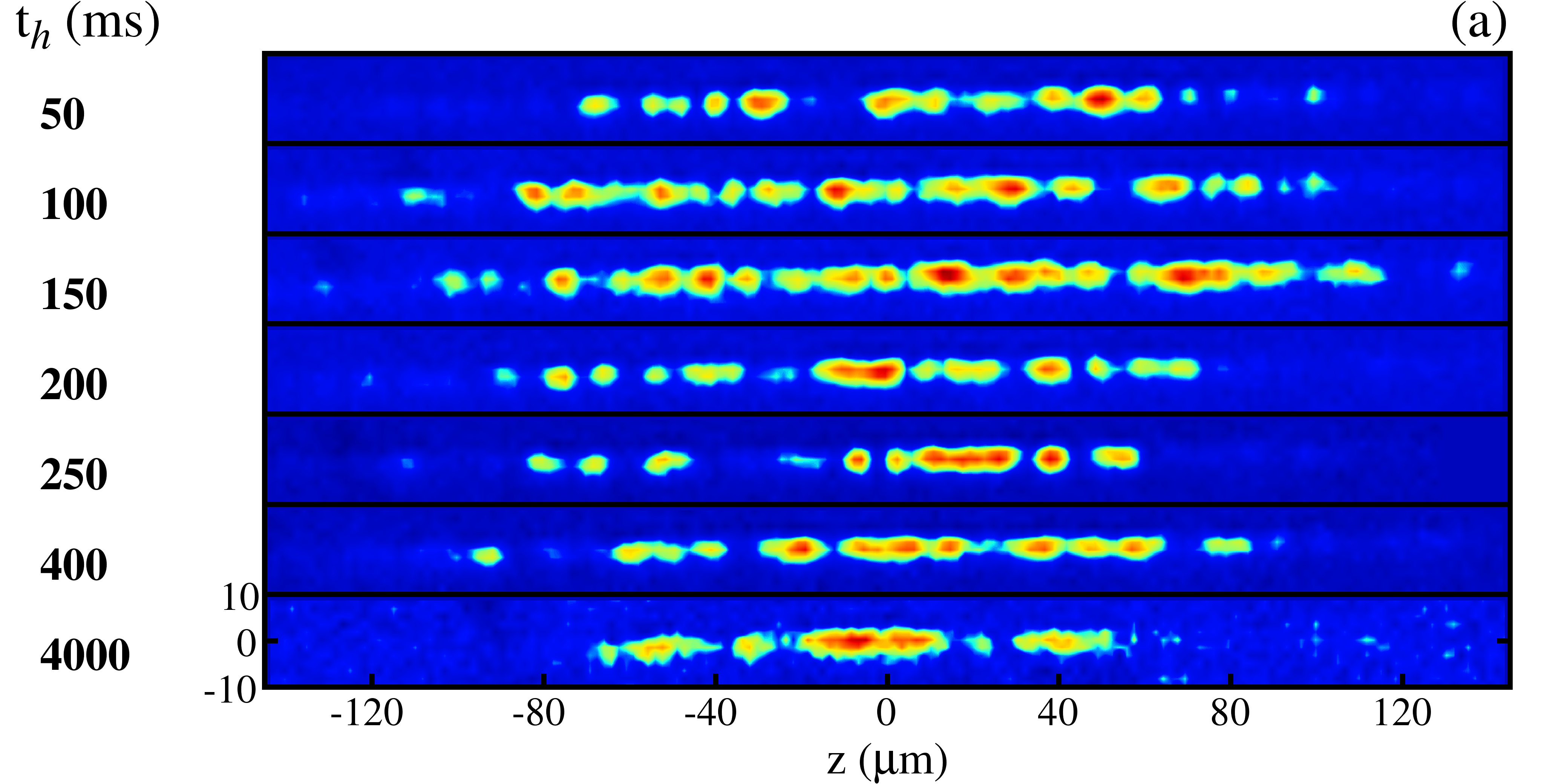} 
\includegraphics[trim={15cm 0 0 0},clip, width=0.42\textwidth]{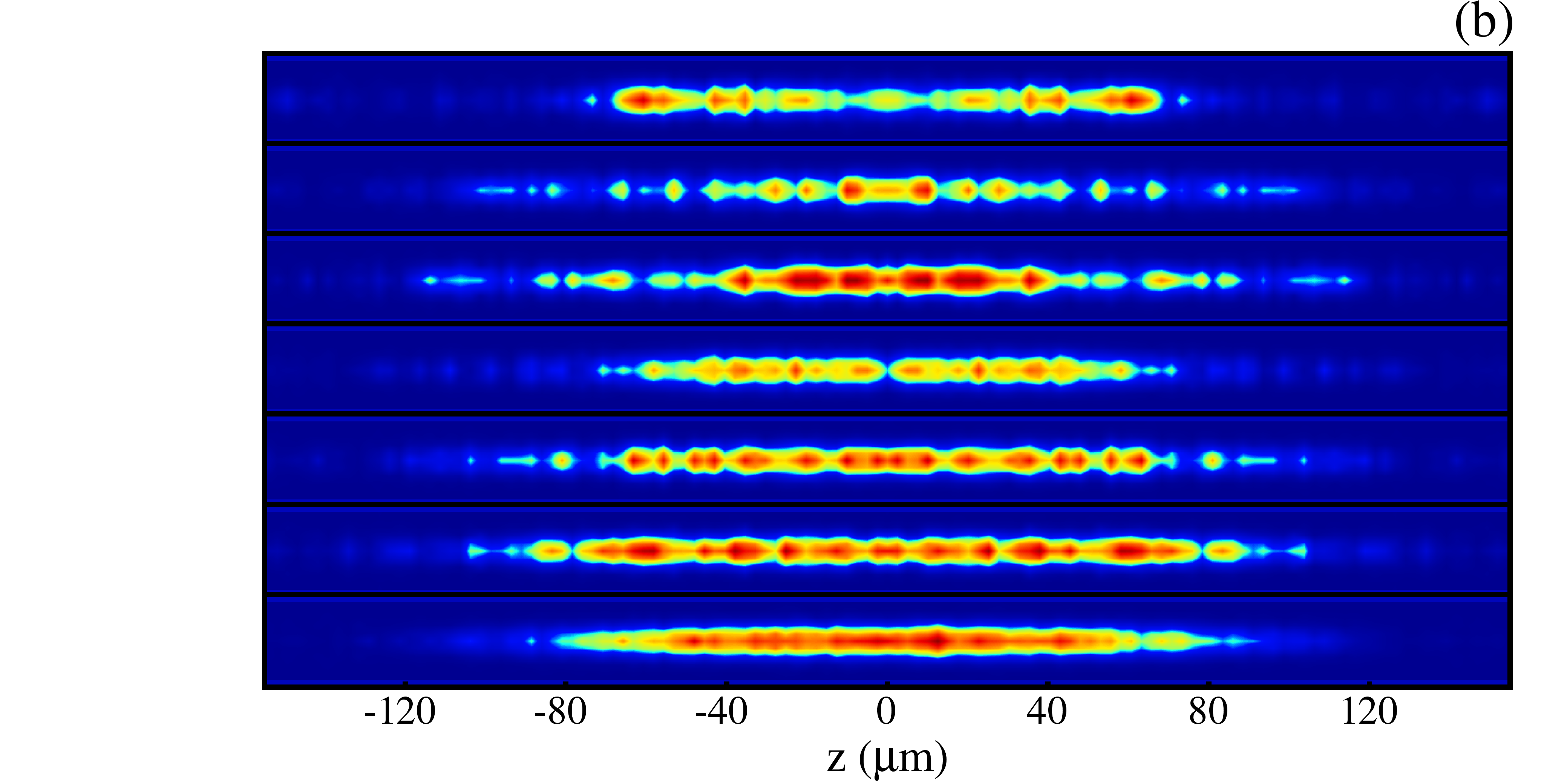}
\caption{(a) Experimental images and (b) GP simulations of column density images for several values of $t_h$, and with $\omega = (2\pi)70\ \mathrm{Hz}$, and $t_m = 250\ \mathrm{ms}$. The axial and radial trap frequencies for the experiments and simulations in this section are $\omega_z = (2\pi)8\ \mathrm{Hz}$ and $\omega_r = (2\pi)254\ \mathrm{Hz}$, respectively. (a) For the experiment, $\bar{B} = 577.4\ \mathrm{G}$ and $\Delta B = 41.3\ \mathrm{G}$, corresponding to $\bar{a} = 5 a_0$, $a_+ = 15 a_0$ and $a_- = -1 a_0$.  Each image, with indicated $t_h$, is a separate realization of the experiment.  (b) Cylindrically symmetric $3$D GP simulations where the calculated $3$D densities are integrated along one transverse direction to produce $2$D column densities.  For the simulations, $a_+ = 20 a_0$ and $a_- = 0.5 a_0$. } 
\label{fig:fig6}
\end{figure}

In Fig.~\ref{fig:fig6}(a) we show experimental images and compare them to Gross-Pitaevskii (GP) simulations.  Note also that the axial and radial trap frequencies in this section are $\omega_z = (2\pi)8\ \mathrm{Hz}$ and $\omega_r = (2\pi)254\ \mathrm{Hz}$, respectively.  We observe that granulation is remarkably persistent in time after the modulation is turned off, and that its structure is random between different experimental runs.  GP simulations for similar parameters are shown in Fig.~\ref{fig:fig6}(b).  In contrast to the experimental images, the GP simulations produce column density distributions that resemble Faraday waves, with a regularly spaced pattern.  Without a stochastic component the GP model represents a crude approximation.  The qualitative difference between the observations in Fig.~\ref{fig:fig6}(a) and the GP simulations in Fig.~\ref{fig:fig6}(b) suggest that the observed state of the atoms in the experiment goes beyond what the GP mean-field theory can describe.

The GP ansatz is a product of one single-particle state $\phi_{GP}$: $\Psi_{GP} \sim \prod_{k=1}^{N}\phi_{GP}(r_k)$.  This is a ``mean-field state'' because all particles in the many-body system occupy the single-particle state $\phi_{GP}(r)$.  A GP product state cannot describe correlations, where the properties of one or several particles in the many-body system depend on the properties of other particles in it.  We go beyond the mean-field GP theory by employing the multiconfigurational time-dependent Hartree for bosons method (MCTDHB or MB), which can account for many-body correlations.   The MCTDHB ansatz incorporates all possible configurations $(n_1, ..., n_M)$ of $N$ particles in $M$ single-particle states, $\vert \Psi \rangle = \sum_{n_1, n_2, ..., n_M} C_{n_1, n_2, ..., n_M} \vert n_1, ..., n_M\rangle$.  The MCTDHB ansatz can therefore self-consistently describe correlations in the many-body state~\cite{SupplMat}.  

We simulate the \emph{in-situ} single-shot images \cite{Sakmann2016,Lode2017} from the wavefunctions obtained with MCTDHB for the various experimental parameters and for $M=2$ modes (see Supplemental Materials \cite{SupplMat} and Refs. \cite{Lode2016b,Fasshauer2016,ultracold.org,Brezinova2012,Sakmann2008,Penrose1956,Spekkens1999,Bouchoule2012,Roati2008} therein).   The simulated single-shot images correspond to drawing random samples from the $N$-particle density $\vert \Psi(r_1,..., r_N) \vert^2$ of the many-body state.  Single-shot images thus contains information about quantum fluctuations and correlation functions of all orders, and the average of many such single-shot images corresponds to the density.   Due to computational constraints, at present, we can only perform 1D simulations.  Along the axial direction, the experimental data show grains that are typically $4-10$ $\mu$m in length in the axial direction while granulation is suppressed transversely, thus justifying the validity of the $1$D approximation and our comparison of $1$D theory with the experimental line densities.

The simulation of single-shot images requires a model of the many-body probability distribution $\vert\Psi(r_1, ..., r_N)\vert^2$ as provided by MCTDHB.  Classical field methods, in contrast, approximate the time-evolution of expectation values using ``clasical-field trajectories'', i.e., solutions of the GP equation with stochastic initial conditions.  These classical-field methods, however, do not supply a model for the wavefunction $\vert\Psi(r_1, ..., r_n)\vert^2$ from which single-shots can be simulated~\cite{Sakmann2016}.  

\begin{figure}
%\vspace*{-10mm}\includegraphics[width=1.0\textwidth]{./figs/figure7.pdf}
\vspace*{-10mm}\includegraphics[width=1.0\textwidth]{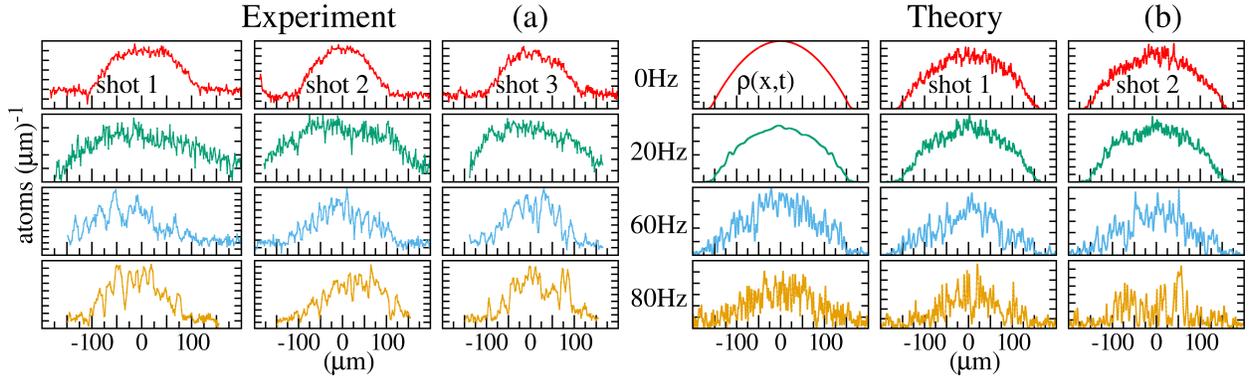}
\vspace*{-00mm}
\caption{Experimental and theoretical line density profiles. (a) Experimental data and (b) many-body simulations for different modulation frequencies.  (a) The rows show data for three independent experimental images (``shots'') for the indicated $\omega$, where $\omega = 0$ corresponds to no modulation.  Here, $\bar{B} = 590.8\  \mathrm{G}, \Delta B = 41.3\ \mathrm{G}$, corresponding to $\bar{a}=8a_0$, $a_+ = 20a_0$, $a_-=0.7a_0$, and $t_m = t_h = 250\ \mathrm{ms}$. (b) The first column shows the density $\rho(z,t)$ as calculated from the $1$D MB theory (see Supplemental Materials) while the second and third columns display two simulated single shots.  We observe that granulation is present in single-shot images, but absent in the average, $\rho(x,t)$.}
\label{fig:fig7}
\end{figure}

Figure \ref{fig:fig7}(a) shows the line density for three independent experimental shots, and for four modulation frequencies, $\omega/2\pi = 0, 20, 60,$ and $80\ \mathrm{Hz}$, where $\omega = 0$ corresponds to no modulation.  For this data the time scales, $t_m = t_h = 250\ \mathrm{ms}$, are much longer than for the data discussed in the context of Faraday waves.  The $1$D MB simulations of the density and, for comparison to experiment, two single shots are shown in Fig.~\ref{fig:fig7}(b).  The single-shot simulations and experimental images are qualitatively similar, in contrast to the densities $\rho(x,t)$, obtained from the MB model.  The shot-to-shot fluctuations in the single-shot simulations result from the fact that single-shots are random samples distributed according to the many-body probability distribution $\vert \Psi(r_1, ..., r_N;t)\vert^2$.  At $\omega = (2\pi) 20\ \mathrm{Hz}$, the experimental line density is somewhat broadened, perhaps indicating an excitation of low-lying quadrupolar oscillations.  For $60\ \mathrm{Hz}$ modulation the single-shot images exhibit large minima and maxima, which are even more pronounced at $80\ \mathrm{Hz}$.  Thus, we find that there is a threshold modulation frequency $\omega_c$, above which the line density is significantly altered.  The density, corresponding to the average of a large number of single shots, does not exhibit grains; they only emerge in single shot images.  

\begin{figure}
%\vspace*{-10mm}\includegraphics[width=0.95\textwidth]{./figs/figure8.pdf}
\vspace*{-10mm}\includegraphics[width=0.95\textwidth]{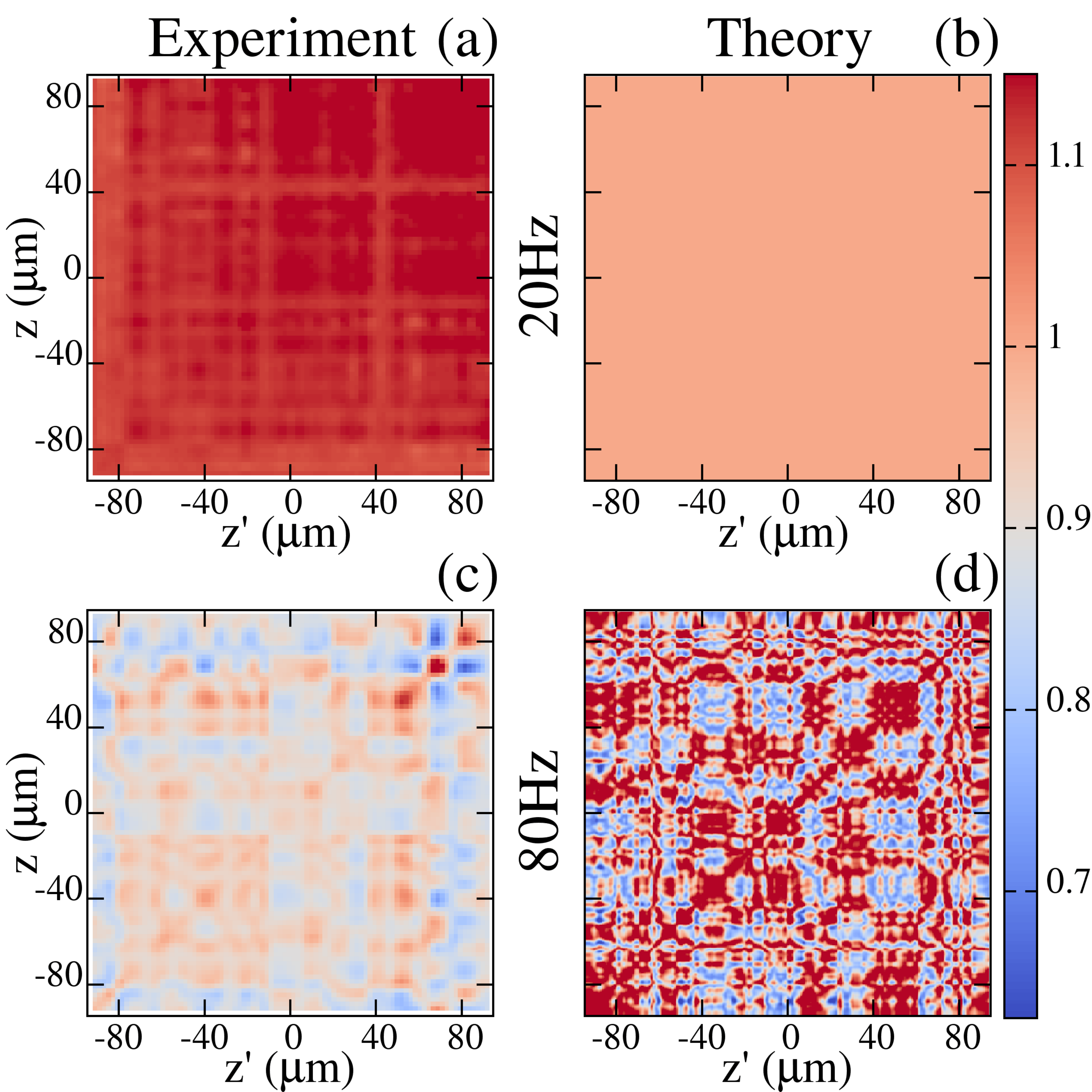}
\vspace*{-00mm}
\caption{$2^{nd}$ order correlation functions.  (a)  Correlation function $C^{(2)}(z,z^{\prime})$  calculated from the experimental data for $\omega = (2\pi) 20\ \mathrm{Hz}$. (b) Correlation function $g^{(2)}(z,z^{\prime})$ calculated from MB theory for the same parameters as (a).  (c) $C^{(2)}(z,z^{\prime})$  calculated from the experimental data for $\omega = (2\pi) 80\ \mathrm{Hz}$. (d) $g^{(2)}(z,z^{\prime})$ calculated from MB theory for the same parameters as (c).   For the non-granulated states ((a) and (b)), the correlation function is ${\sim}1$, indicating the absence of $2^{nd}$ order correlations.  For the granulated states ((c) and (d)), regions with correlations (red hues) and anti-correlations (blue hues) emerge.  Theoretical and experimental $2^{nd}$ order correlations qualitatively agree: they are flat for the non-granular states ((a) and (b)) and exhibit patterns of comparable length-scale and magnitude for granular states ((c) and (d)).  All images correspond to $t_{h}=t_{m}=250\ \mathrm{ms}$ and $\bar{a} = 8a_0$, $a_{+}=20$, and $a_{-}=0.5a_0$.}
\label{fig:fig8}
\end{figure}

Figure~\ref{fig:fig8} shows the $2^{nd}$ order correlation functions for the experiment $C^{(2)}(z,z^{\prime})$, and MB theory $g^{(2)}(z,z^{\prime})$, where both quantities are defined in the Supplementary Materials~\cite{SupplMat}.    $C^{(2)}(z,z^{\prime})$ are evaluated using an average of up to $4$ experimental shots, whereas $g^{(2)}(z,z^{\prime})$ are computed directly from the MCTDHB wavefunctions.  

In both the experiment (Fig.~\ref{fig:fig8}(a)) and MB theory (Fig.~\ref{fig:fig8}(b)), we find that when $\omega < \omega_c$ the condensate is practically uncorrelated, as evidenced by $C^{(2)}(z,z^{\prime})\approx g^{(2)}(z,z^{\prime}) \approx 1$.  However, when $\omega > \omega_c$ we find that the relatively constant correlation plane evolves into smaller correlated and anti-correlated regions, as shown Fig.~\ref{fig:fig8}(c) for the experiment and Fig.~\ref{fig:fig8}(d) for the MB theory.

\begin{figure}
\includegraphics[width=1\textwidth,angle=-90]{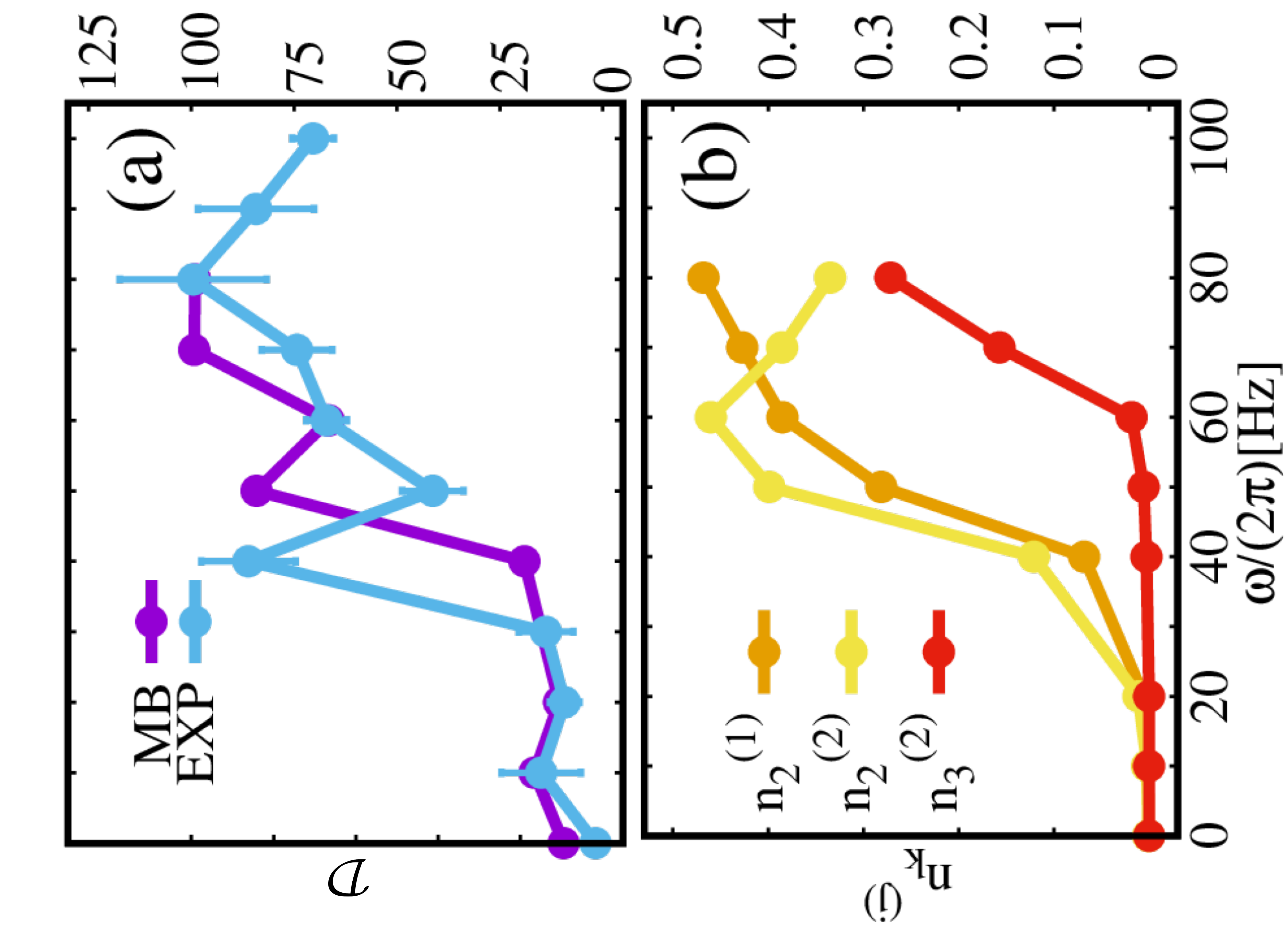}
\caption{Granulation vs. $\omega$. (a) Comparison of the deviations from a Thomas-Fermi distribution as quantified by the contrast parameter $\mathcal D=\mathcal D(\omega)$~\cite{SupplMat} for single shots simulated with the MB theory with those taken in experiment (EXP). MB theory predicts the threshold value, $\omega_c \approx (2\pi) 30\ \mathrm{Hz}$, where deviations become large and grains form. Each symbol and its error bar are the mean and standard error of the mean of at least $4$ experimental measurements of $\mathcal D$, while $100$ single shots at each $\omega$ have been used for the MB simulations. (b) Eigenvalues of the first and second order RDM.  A growth of all three are observed to occur for $\omega > \omega_c$, indicating the emergence of correlations and fragmentation.  The growth of both $n_2^{(1)}$ and $n_2^{(2)}$ occur as $\omega \approx \omega_c$, with the drop in $n_2^{(2)}$ near $60\ \mathrm{Hz}$ corresponding to the subsequent growth in $n_3^{(2)}$.}
\label{fig:fig9}
\end{figure}

To further characterize the granulated states, we plot the contrast parameter $\mathcal{D}$ at each modulation frequency in Fig~\ref{fig:fig9}(a).  $\mathcal{D}$ quantifies the deviation of a given set of single shots from a parabolic fit -- as discussed in the Supplementary Materials~\cite{SupplMat} and Fig.~S1 therein.  A sharp threshold can be seen both in the experimental data and the simulations at $\omega_c \approx (2\pi) 30\ \mathrm{Hz}$, beyond which grains start to form.  For $\omega<\omega_c$ the gas oscillates coherently without significant deviation from a Thomas-Fermi envelope.

% Figure 3

The threshold frequency $\omega_c$, can be understood by examining the 2$^{nd}$ largest eigenvalues (or occupations) $n_2^{(1)}$ and $n_2^{(2)}$ of the 1$^{st}$ and 2$^{nd}$ order reduced density matrices (RDMs), respectively (see Supplemental Materials \cite{SupplMat}), which are plotted in Fig.~\ref{fig:fig9}(b).  These may be used as a measure of the departure of our MB model from mean-field states.  Many-body systems, where multiple eigenvalues of the 1$^{st}$ order RDM are macroscopic (ie.~of order $N$), are termed \emph{fragmented}~\cite{Spekkens1999,Mueller2006}.  At zero excitation only $n_1^{(1),(2)}$ are macroscopic while $n_2^{(1),(2)}$ are nearly zero.  The latter increase substantially with $\omega$ beyond $\omega_c$, heralding the loss of 1$^{st}$ and 2$^{nd}$ order coherence and the emergence of correlations as shown in Fig.~\ref{fig:fig8}.  At $\omega \approx (2\pi) 50 \ \mathrm{Hz}$ we observe a drop in $n_2^{(2)}$, however, this results in an increase in $n_3^{(2)}$ and not an increase in $n_1^{(1),(2)}$.  The MCTDHB computations thus show that the emergence of granulation is accompanied by the conversion of initial condensation (only a single macroscopic occupation~\cite{Sakmann2008}) into fragmentation.

Both observations, the emergence of fragmentation and the loss of 2$^{nd}$ order coherence, underscore that the granulation of Bose-Einstein condensates is a many-body effect. The system thus cannot be described by a mean-field product state any longer and has left the realm of GP theory. Although the transition to fragmentation is not sharp -- the natural occupations $n_i^{(1),(2)}$ take on continuous values -- it is well established at sufficiently large $\omega$.  Granulation features randomly-distributed variably-sized grains of atoms which can be observed in single shot images.  Fragmentation, or depletion, on the other hand, is characterized by the reduced density matrix and its (macroscopic) eigenvalues and is not necessarily accompanied by granulation of the density~\cite{Spekkens1999,Mueller2006}.  In our close-to-one-dimensional setup, we observe granulation to emerge side-by-side with fragmentation.

\begin{figure}
\includegraphics[width=0.85\textwidth,angle=-90]{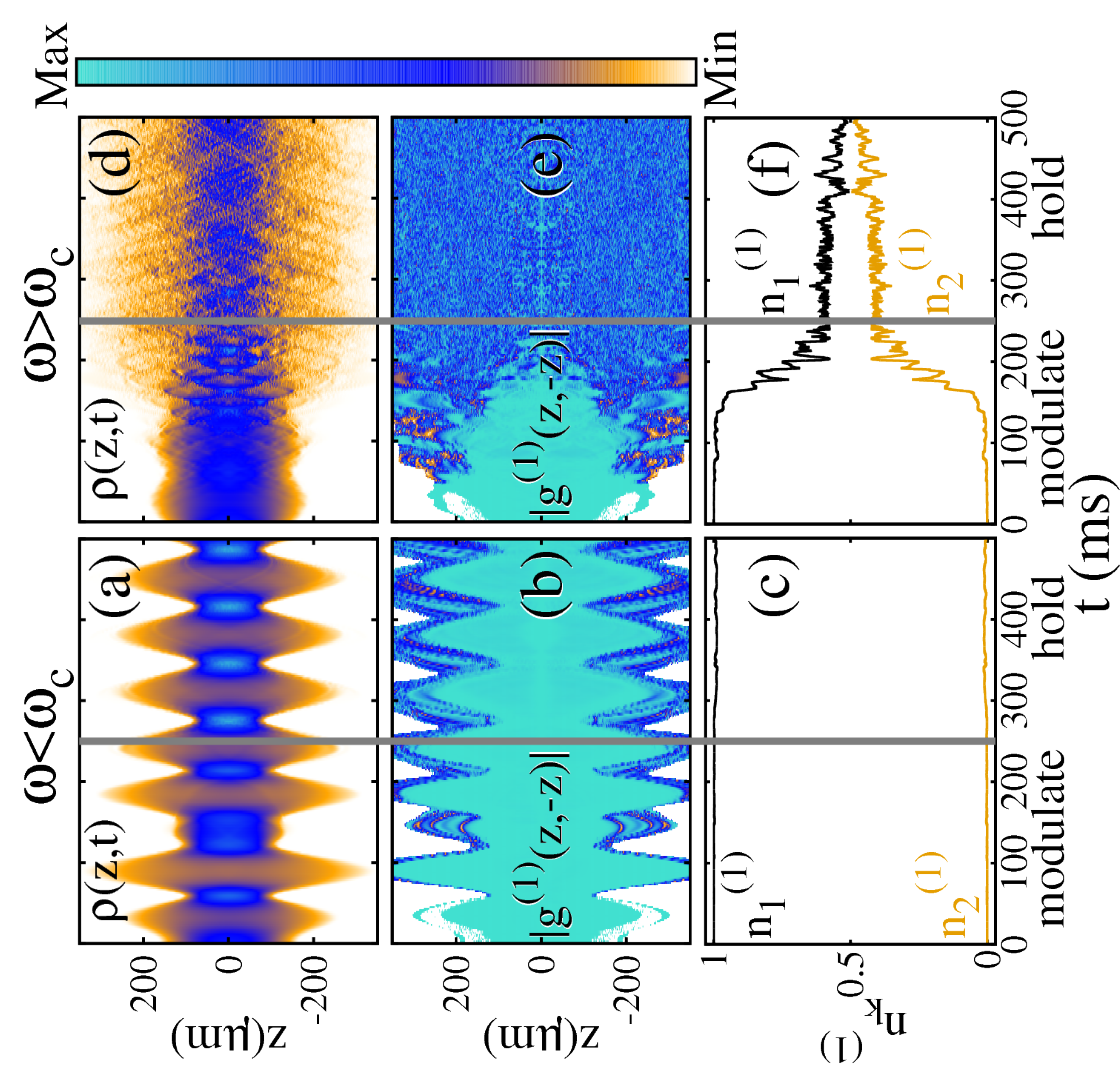}
\caption{Time-evolution, coherence, and fragmentation from simulations. (a,d) The density $\rho(z,t)$, (b,e) first-order spatial correlation function $|g^{(1)}(z,-z)|$; and (c,f) natural occupations $n^{(1)}_k(t)$ are plotted vs.~time $t$. $n_1^{(1)}$ is denoted by the black line, while $n_2^{(1)}$ is indicated by the yellow line,  Panels (a)--(c) are calculated with $\omega = (2\pi)20\ \mathrm{Hz}< \omega_c$ and panels (d)--(f) with $\omega = (2\pi)80\ \mathrm{Hz}> \omega_c$. All other parameters are given in the Fig.~\ref{fig:fig7} caption. The onset and formation of granulation as inferred by the simultaneous drop in the values of $|g^{(1)}|$ and $n^{(1)}_1$, indicating the emergence of spatial correlations and fragmentation, respectively. }
\label{fig:fig10}
\end{figure}

The dynamical evolution, as calculated from the MB theory, of the density is shown in Fig.~\ref{fig:fig10}(a) and \ref{fig:fig10}(d) for $\omega < \omega_c$ and $\omega > \omega_c$, respectively.  In both cases, the modulation of the Thomas-Fermi radius follows the external perturbation.  Once the modulation is turned off, the radius oscillates at its natural quadrupolar frequency.  The $1^{st}$-order spatial coherence is shown in Fig.~\ref{fig:fig10}(b,e) for the same parameters.   The patterns that emerge and persist in $g^{(1)}(z,z^{\prime})$ demonstrate that spatial correlations between particles at distinct and distant locations in the granular state are present [Fig.~\ref{fig:fig10}(e)].  The length-scale of the patterns in $g^{(1)}(z,z^{\prime})$ is similar to what is seen in Fig.~\ref{fig:fig8} for $g^{(2)}(z,z^{\prime})$.  We infer that the process of granulation in a BEC is accompanied by the emergence of non-local correlations in the many-body state.  Fig.~\ref{fig:fig10}(f) shows the emergence of two macroscopic eigenvalues of the reduced one-body density matrix for $\omega > \omega_c$. While these so-called natural occupations are unaffected by modulation for $\omega < \omega_c$, as seen in Fig.~\ref{fig:fig10}(c), $\omega > \omega_c$ results in the second natural orbital being macroscopically populated, and hence, in the fragmentation of the BEC [Fig.~\ref{fig:fig10}(f)].  An examination of the total energy per particle ($E_t$) imparted during modulation for a time $t_m$ shows that $E_t \approx 22\ \mathrm{nK}$ when $\omega = (2\pi) 20 \ \mathrm{Hz}$, and $E_t \approx 36\ \mathrm{nK}$ when $\omega = (2\pi) 80\ \mathrm{Hz}$, both of which are much less than the critical temperature $T_c\approx 330\ \mathrm{nK}$.

\begin{figure}
\includegraphics[width=1\textwidth]{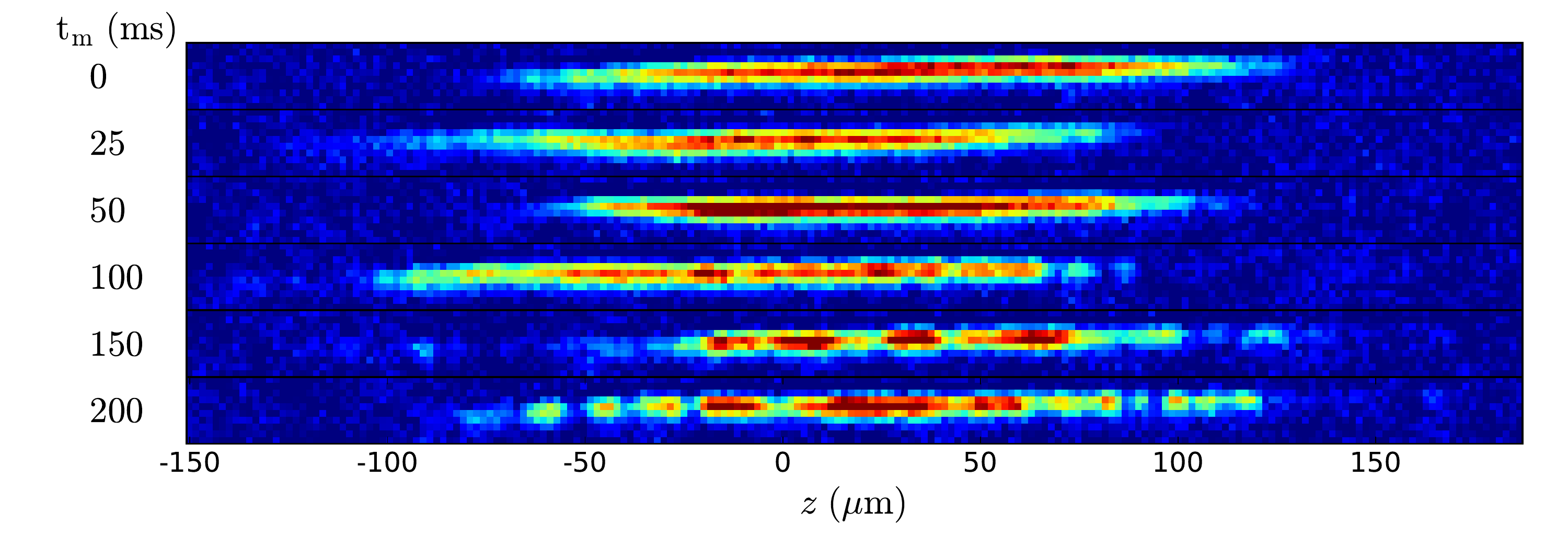}
\caption{Experimental column densities showing the formation of grains.  Representative column density images taken at different $t_m$.  For each value of $t_m$, $\omega=(2\pi) 70\ \mathrm{Hz}$ and $t_h = 250\ \mathrm{ms}$.  All other parameters are given in the Fig.~\ref{fig:fig7} caption.  Each image is a different realization of the experiment.}
\label{fig:fig11}
\end{figure}

The onset of granulation observed experimentally is shown in Fig.~\ref{fig:fig11}.  The condensate was modulated at $\omega= (2\pi) 70 \ \mathrm{Hz}$ for various $t_m$ followed by $t_h=250\ \mathrm{ms}$.  For $t_m < 100\ \mathrm{ms}$ there is no discernable difference between the modulated and unmodulated ($t_m = 0 \ \mathrm{ms}$) cases, but for $t_m > 100\ \mathrm{ms}$ grains are observed to form.  Consistent with Fig.~\ref{fig:fig10}, the transition to a granulated state is gradual with increasing $t_m$.  The observed grains are also long-lived in comparison to Faraday waves, as shown in Fig.~\ref{fig:fig6}(a) and Fig.~\ref{fig:fig5}(a), respectively.

The transition to granular states occurs due to the presence of quantum correlations.  The $2^{nd}$ order correlations, shown in Fig.~\ref{fig:fig8}, and the $1^{st}$ order, non-local correlations, shown in the middle panel of Fig.~\ref{fig:fig10} result from modulating the interaction and do not disappear after the modulation is stopped.   Our modeling of the state on the many-body level suggests that granulation represents a dynamical many-body state characterized by the presence of quantum fluctuations, correlations, fragmentation, and their persistence in time.

Granular states feature random patterns and lack periodicity in their distributions, distinguishing them from Faraday and shock waves~\cite{Perez2004}. The multi-characteristic nature of quantum grains is supported by our observation of additional anomalous features in real and momentum space. Indeed, we find signatures of different co-existing phases of perturbed quantum systems such as quantum turbulence and localization in granulated states. We verified that the density in momentum space (as calculated from the MB theory) of the granulated state shows clear signs of a $k^{-2}$ power-law scaling (see Supplemental Materials \cite{SupplMat} and Fig.~S2 therein) which indicates a connection to turbulent BECs~\cite{Thompson2014,Navon2016,Tsatsos2016}.

\section{Conclusions}

We have explored the response of a BEC to modulated interactions.  In the regime where the drive frequency $\omega \gtrsim \omega_r$, the drive couples to parametric and resonant modes that result in $1$D spatial pattern formation.  For $\omega$ near resonant with $2\omega_r$ or $\omega_r$, very little modulation time and amplitude are required to produce a significant response.  Near these resonances the condensate undergoes breathing oscillations that persist for a long time, resulting in the formation of Faraday and resonant mode patterns for $t_h >0$.  A pattern is also observed off-resonance, but only with increased modulation amplitude and modulation time.  Due to the long modulation time, the resulting pattern can be seen at $t_h=0$, and is a direct consequence of the applied modulation.  The dispersion relation of both Faraday and resonant modes is well-represented by a mean-field theory that accounts for the 3D nature of the elongated condensate. 

For lower drive frequencies, the modulated interactions only weakly couple to the condensate.  Significant response is achieved only by increasing the modulation amplitude and time, and then, only above a critical modulation frequency $\omega_c$.  Fluctuating and irregular spatial patterns, that we define as grains, may then emerge and persist for long periods of time.  A theoretical description of granulation requires approaches that go beyond mean-field theory, indicating that quantum granulation is characterized by non-local many-body correlations and quantum fluctuations.

\begin{acknowledgments}
This work was supported in part by the Army Research Office Multidisciplinary University Research Initiative (Grant No. W911NF-14-1-0003), the Office of Naval Research, the NSF (Grant No. PHY-1707992), the Welch Foundation (Grant No. C-1133), the Austrian Science Foundation (FWF) under grant No. F41(SFB `ViCoM') and No. P32033, the Wiener Wissenschafts- und TechnologieFonds (WWTF) project No. MA16-066 (`SEQUEX') and by FAPESP, under CEPID program (Grant No. 2013/07276-1).  Computational time in the High-Performance Computing Center Stuttgart (HLRS) is gratefully acknowledged.  We also thank Mustafa Amin for valuable discussions.
\end{acknowledgments}

\newpage

\begin{center}
\Large{Supplementary Materials for} \\
\Large{Parametric Excitation of a Bose-Einstein Condensate:  From Faraday Waves to Granulation} \\
\large{J. H. V. Nguyen, M. C. Tsatsos, D. Luo, A. U. J. Lode, G. D. Telles , V. S. Bagnato, R. G. Hulet}
\end{center}

%~\\~\\~\\
%\textbf{This PDF includes:} \\
%Materials and Methods \\
%Supplementary Text \\
%Fig. S1 \\
%Fig. S2 \\
%Fig. S3 \\ 
%Fig. S4 \\
%Fig. S5 \\
%Fig. S6 \\
%Fig. S7

\section*{Experimental Details}

A pair of coils in Helmholtz configuration is used to produce a homogeneous magnetic field, $B$, which allows us to vary the interatomic interactions. For a given value of $B$ the corresponding scattering length is determined from:

\begin{equation}
a= a_{\rm{bg}}\left(1+\frac{\Delta}{B-B_{\infty}}\right),
\label{eq:aofB}
\end{equation}
where $a_{\rm{bg}}=-24.5\ a_0$, $B_{\infty}=736.8\ \mathrm{G}$, $\Delta=192.3\ \mathrm{G}$ \cite{Pollack2009} and $a_0$ the Bohr radius.  An oscillation of the bias field, $B(t)=\bar{B} + \Delta B sin(\omega{}t)$, where $\bar{B}$ is the mean and $\Delta {B}$ is the modulation amplitude.  This produces an asymmetric $a(t)$ since $a$ is a non-linear function.  Thus, the mean is $\bar{a}$, the maximum is $a_+$, and minimum is $a_-$.

\section*{Numerical Method: MCTDHB}
The Hamiltonian describing the problem is:
\begin{equation}
\mathcal H(t) = \mathcal T + \mathcal V + \mathcal W(t),
\label{hamiltonian}
\end{equation}
with $\mathcal T=-\frac{\hbar^2}{2m} \sum_i^N \nabla_{\bf r_i}^2$, $\mathcal V=\sum_i^N V_{\rm{trap}}(\bf r_i)$ and $\mathcal W=\sum_{i<j}W(\mathbf r_i-\mathbf r_j;t)$ being the many-body kinetic, potential, and interaction energy operators, respectively. We have:
\begin{eqnarray}
V_{\rm{trap}}(\mathbf r) &=& \frac{\omega_z^2}{2}z^2 + \frac{\omega_r^2}{2}r^2 ~~~~\rm{and} \\
W({\mathbf r}_i-{\mathbf r}_j;t) &=& g(t) \delta(|{\mathbf r}_i - {\mathbf r}_j|) = g_0 \left[-\beta_1 + \frac{\beta_1}{\beta_2-\beta_3\sin(\omega t)}\right] \delta(|{\mathbf r}_i - {\mathbf r}_j|),
\label{Hamiltonian}
\end{eqnarray}
where $g(t)$ and $g_0$  are dimensionless parameters quantifying the time-dependent and time-independent interaction strengths, respectively, whose values are given below, $\beta_1=-\beta_2/(\beta_2-1)=|a_{bg}/\bar{a}|=24.5/7.9$, $\beta_2=|(\bar{B}-B_{\infty})/\Delta|$, $\beta_3=|\Delta B /\Delta|$, and $\mathbf{r}=(x,y,z)^T$. 
The time-dependent interparticle interaction models the experimental modulation of the scattering length. In the granulation experiment $\omega_r/\omega_z\approx 32$ and so the trap has a cigar shape, close to the $1$D regime~\cite{Menotti2002}. 

To solve the time-dependent Schr\"odinger equation for many interacting particles,
\begin{equation}
i \hbar \frac{\partial \Psi}{\partial t} = \mathcal H(t) \Psi \label{TDSE},
\end{equation}
we apply the Multiconfigurational Time-Dependent Hartree theory for Bosons (MCTDHB) \cite{Streltsov2007,Alon2008} and use the MCTDH-X numerical solver \cite{Lode2016b,Fasshauer2016,ultracold.org} for $1$D and $3$D simulations. The MCTDHB theory assumes a general ansatz $\Psi=\Psi(\mathbf R,t)$ for the $N-$particle problem and expands it on a many-body basis $\Psi({\mathbf R},t)=\sum_k C_k(t) \Phi_k({\mathbf R}, t)$, where $\Phi_k$ are all possible permanents (i.e. boson-symmetrized many-particle wavefunctions) built over a finite set of $M$ orbitals (i.e. single-particle orthonormal states) $\phi_j(\mathbf{r})$ and ${\mathbf R}=\{\mathbf{r_1},\mathbf{r_2},\dots,\mathbf{r_N}\}$. The theory goes beyond the standard mean-field approximation and incorporates fragmentation and correlation functions of any order $p$, $1\le p\le N$~\cite{Sakmann2016}. Note that the $M$ orbitals are found self-consistently and are \emph{not a priori} chosen. Therefore, MCTDHB chooses the best set of orbitals at each time. We performed three sets of simulations with the following parameters:\begin{enumerate}
\item\label{item1} A one-dimensional system, with (dimensionless) trap frequency $\omega_{\rm{com}}=0.1$, $N=10^4$, $M=1$ and $M=2$ and $g^{(1D)}=g_0(N-1)=357$. The interaction parameter is found from $g^{(1D)}=2 a N_{\rm{exp}} \sqrt{\omega_{\rm{com}}} l_z/l_r^2$, where $a$ is the experimental background value of the scattering length and $l_{r,z}=\sqrt{\hbar/(m\omega_{r,z})}$. The experimental trap frequencies $\omega_r=(2\pi)254$ Hz, $\omega_z=(2\pi)8$ Hz  have been used  and $\bar{a}=7.9 a_0$, $N_{\rm{exp}}=5.7\times 10^5$ particles (see Eq.~\ref{Hamiltonian}). 
The simulations and quantities derived from this dataset are presented in Figs.~7--10. The computation was performed on a $1$D spatial grid of 4096 points. 
The modulating frequencies take on the values $\omega /(2\pi)=10,20,30,\dots,90$ Hz. Due to the fast temporal modulation of the atom-atom interaction operator and the resulting strong local density modulations, the computations are numerically highly demanding.  Therefore, extended convergence checks are required. We have confirmed convergence with respect to both the spatial grid density and the integration time step as well as error tolerance for frequencies up to $\omega/\omega_z=10$. Even though the error tolerance demanded is $10^{-11} - 10^{-10}$ (extremely high accuracy) the accumulated error in the total energy at the end of the propagation remains between $3-8\%$ and is somewhat larger for the natural occupations. This reflects the fact that the Fock space (spanned by $M=2$ basis functions) is far from complete~\cite{Brezinova2012}.

At $\omega/2\pi=30\ \mathrm{Hz}$ we have seen resonant behavior: the energy increases up to $\approx 10$ times after $500\ \mathrm{ms}$ and the density is found to occupy all available space. Convergence checks are beyond the computational capacities and the point at $\omega/2\pi=30\ \mathrm{Hz}$ has not been included in the plots. We attribute the resonant behavior at $\omega/2\pi=30\ \mathrm{Hz}$ to its proximity to $2\omega_Q$, where $\omega_Q=\sqrt{3}\omega_z$ is the 1D quadrupolar frequency.  We have also performed calculations for $\omega_Q=13.9\ \mathrm{Hz}$ and $2\omega_Q=27.9\ \mathrm{Hz} $, and have observed similar behavior. 
%
%
%\item \label{item2} A one-dimensional set of systems with $\omega_z=\sqrt{0.1}$, $g_0(N-1)=357$ for $i)$ variable number of orbitals $2\leq M\leq 4$ and $N=100$ and $ii)$ variable number of particles $100\leq N\leq 50,000$ and $M=2$ (Fig.~S2). The rest of the parameters are identical to the dataset of paragraph \ref{item1} (see above). In all cases it was found that fragmentation persists both as a function of $M$ and $N$.
%
%
\item\label{item2} A three-dimensional system with $\omega_z=1, \omega_y=\omega_x=32$, $N=1000$, $M=1$ and $g^{(3D)}=4\pi N_{\rm{exp}} a/l_z = 222$, using also a delta-type interaction pseudopotential. The computational grid was $512\times{}64\times{}64$ wide. All other parameters are set as in paragraph \ref{item1}. The modulation frequency was set to $\omega=8.75\omega_z$, that corresponds to the experimental value $\omega/2\pi=70$Hz. The amplitude of modulation of the interaction is, as before, always positive (results plotted in Fig.~6). 
\end{enumerate}
All our simulations use a discrete variable representation. The orbital part of the MCTDHB equations of motion are solved using Runge-Kutta or Adams-Bashforth-Moulton of fixed order (between 5 and 8) and variable stepsize as well as the Bulirsch-Stoer scheme of variable order and stepsize. Davidson diagonalization  %\cite{Payne_RevModPhys.64.1045} 
and short iterative Lanczos schemes %\cite{Lanczos_J.Res.Natl.Bur.Std.45.255} 
were used to evaluate the coefficient part of the MCTDHB equations. The stationary initial state $\Psi_0$ is found by imaginary time propagation with time-independent interactions, $g(t)=g_0$. Subsequently, $\Psi_0$ is propagated in real time for the above time-dependent Hamiltonian and sets of parameter values. 
For the parameters chosen in the $3$D simulation the time unit is $\tau=19.9$ms and the length unit is $L=13.5\mu$m. For the $1$D simulations we have $\tau=2$ms and $L=4.3\mu$m. Energy is measured in units of $\hbar^2/(mL^2)$.

\section*{Correlation functions, single shot simulations, and contrast}

The density matrix $\rho^{(N)}=|\Psi\rangle\langle\Psi|$ describes the $N$-body quantum system in state $\Psi$ and the reduced density matrix (RDM) of order $p=1,2\dots$ (partial trace of $\rho^{(N)}$) is most commonly employed and gives the $p$-particle probability densities. The eigenbasis of the RDMs gives information on the $p^{th}$-order coherence of the system. In particular, if there is more than one macroscopic eigenvalues of the first (second) order RDM then the system is fragmented and first (second) order coherence is lost.

Specifically, the $p^{th}$-order reduced density matrix (RDM) is defined as \cite{Sakmann2008}:
\begin{eqnarray}
\label{RDM}
&\rho^{(p)}(z_1,\dots,z_p|z_1^\prime,\dots,z_p^\prime;t) \nonumber \\
&= \frac{N!}{(N-p)!} \int \Psi(z_1,\dots,z_p, z_{p+1}, \dots, z_N;t) \\ & \times \Psi^\ast(z_1',\dots,z_p', z_{p+1}, \dots, z_N;t) dz_{p +1}\dots dz_N \nonumber \\
 &=\sum_k n^{(p)}_k(t) \phi^{(p)}_k(z_1,\dots,z_p;t) \phi_k^{(p)\ast}(z_1',\dots,z_p';t), &
\end{eqnarray}
where $n^{(p)}_k(t)$ are its eigenvalues and $\phi^{(p)}_k(t)$ its eigenfunctions. For $p=1$, $n^{(1)}_k(t) \equiv n_k(t)$ are the so-called \emph{natural occupations} of the corresponding \emph{natural orbitals} $\phi^{(1)}_k(z;t)$. According to the Onsager-Penrose definition \cite{Penrose1956}, a system of $N$ interacting bosons is said to be condensed if and only if one natural orbital $\phi^{(1)}_m$ is macroscopically occupied, or, $n_m/N \sim 1$ for some $m$, while $n_j/N\sim 0$ for $j\neq m$. If more than one natural orbital is macroscopically occupied then the system is called \emph{fragmented} \cite{Spekkens1999}.
The diagonal 
\begin{equation}
\rho(z;t)\equiv\rho^{(1)}(z|z;t)=\sum_{k=1}^M n_k(t)|\phi^{(1)}_k(z;t)|^2
\label{EqDensity}
\end{equation}
we simply call \emph{density}.
The eigenfunctions $\phi_{k}^{(2)}(z_1,z_2)$ of the 2$^{nd}$ order RDM are known as \emph{natural geminals} (NG). Their occupations satisfy $\sum_{j=1} n^{(2)}_j =N(N-1)$ and are plotted in Fig.~9(c) (normalized to 1).

The $p$th order correlation function is:
\begin{eqnarray}
\label{corrfun}
g^{(p)}(z_1,\dots,z_p|z_1^\prime,\dots,z_p^\prime;t) = \frac{\rho^{(p)}(z_1,\dots,z_p|z_1^\prime,\dots,z_p^\prime;t)}{\sqrt{\prod_{i=1}^p \rho^{(1)}(z_i,z_i;t) \rho^{(1)}(z_i^\prime,z_i';t)}}.
\end{eqnarray}
The skew diagonal (antidiagonal)
\begin{equation}
\label{corrskew}
g_{skew}(z,t)=g^{(1)}(z,-z;t).
\end{equation}
gives the degree of correlation of the density at a point $z$ with its antipodal at point \cite{Bouchoule2012} $z'\equiv -z$ (see Fig.~9).
Similarly, the normalized $p$th order correlation function in momentum space can be defined,
%\begin{eqnarray}
%\label{corrfunmom}
%g^{(p)}(k_1,\dots,k_p|k_1^\prime,\dots,k_p^\prime;t) = \frac{\rho^{(p)}(k_1,\dots,k_p|k_1^\prime,\dots,k_p^\prime;t)}{\sqrt{\prod_{i=1}^p \rho^{(1)}(k_i,k_i;t) \rho{(1)}(k_i^\prime,k_i';t)}},
%\end{eqnarray}
via the Fourier transform $\tilde\rho^{(p)}(k_1,\dots,k_p|k_1^\prime,\dots,k_p^\prime;t)$ of $\rho^{(p)}(z_1,\dots,z_p|z_1^\prime,\dots,z_p^\prime;t)$.
Note that $|g^{(1)}|$, the spatial correlation function, is bounded like $0\leq|g^{(1)}|\leq1$ for any two points $(z,z')$. For Bose condensed and hence non-fragmented states, $|g^{(1)}|$ takes its maximal value everywhere in space and the state is first-order coherent. Moreover, if $|g^{(2)}|<1$ we term the state \emph{anticorrelated} while for $|g^{(2)}|>1$ we term it \emph{correlated}.

The $2^{nd}$ order correlation function of Fig.~8(a) and Fig.~8(c) for some observed distributions $n(z)$ is given by:
\begin{equation}
C^{(2)}(z,z^\prime) =  \frac{ \left\langle n(z) n(z^\prime) \right\rangle}{ \left\langle n(z) \right\rangle \left\langle n(z^\prime)\right\rangle }.
\label{EqCorF2}
\end{equation}
We emphasize that in the expression for $C^{(2)}$, the notation $\left\langle \cdot \right\rangle$ corresponds to an average across experimental realizations. 

The \emph{single-shot simulations} plotted in Fig.~7(b) and Fig.~S1 have been obtained with the method to obtain random deviates of the $N$-particle probability density $\vert \Psi \vert^2$ that is prescribed in Refs. \cite{Sakmann2016,Lode2017}. In brief, the procedure relies on sampling the many-body probability density $\vert \Psi \vert^2$ as follows: one calculates the density $\rho_0(z)$, from the obtained solution $|\Psi^{(0)}\rangle\equiv \Psi$ of the MCTDHB equations. A random position $z_1'$ is drawn from $\rho_0(z)$. In continuation, one particle is annihilated at $z_1'$, the reduced density $\rho_1$ of the reduced system $|\Psi^{(1)}\rangle$ is calculated and a new random position $z_2'$ is drawn. The procedure continues for $N-1$ steps and the resulting distribution of positions $(z_1',z_2',\dots,z_N')$ is a simulation of an experimental single-shot image.

The \emph{contrast parameter} $\mathcal D$ quantifies the deviation of some spatial distribution $n(z)=n(z;t_0)$ of a single shot at a given time $t_0$ from the parabolic (Thomas-Fermi-like) best fit $n_{\rm{bf}}(z)=n_{\rm{bf}}(z;t_0)$ at the same time and is defined as:
\begin{eqnarray}
\mathcal D &=& \int dz \frac{|n(z)-n_{\rm{bf}}(z)|}{n_{\rm{bf}}(z)} ~~~ \textnormal{or} \\
\mathcal D &=& \sum_i^{n_{\rm{gp}}} \frac{|n(i)-n_{\rm{bf}}(i)|}{n_{\rm{bf}}(i)}, ~~~~~\textnormal{iff}~~~ |n(i)-n_{\rm{bf}}(i)|\geq C_{\rm{cutoff}},
\label{EqContrastF}
\end{eqnarray}
where $i$ runs over all $n_{\rm{gp}}$ pixels/grid points. The cutoff requirement $C_{\rm{cutoff}}=0.20~n_{\rm{bf}}(0)$ is set so that small (zero-excitation) fluctuations are wiped out and only values with large deviations are considered (see Fig.~S1). Therefore, the resulting contrast parameter reflects only the large deviations of a given density from its parabolic best fit. 
%Note that, through the above definition, $\mathcal D$ does not depend neither on the normalization of $n(z)$ nor the number $n_{\rm{gp}}$. 
To determine the best fits we used the {\tt{gnuplot}} software to fit the polynomial $p(z)=-a(z-b)^2+c$, where $a,b,c\in \mathbb R$, to the obtained experimental or numerical distributions $n(z)$ along $z$. The two-dimensional experimental column densities have been integrated along $y$. 
The experimental data were also interpolated to a number of points along $z$ so as to equal the grid used for the numerical simulations. 
An example of a processed image is shown in Fig.~S1.

%%
%%
%%
%% Supplemental Material FIGURES
%% Supplemental Material FIGURES
%% Supplemental Material FIGURES
%%
%%\clearpage
%%\begin{figure}
%\hspace*{00mm}\includegraphics[width=0.65\textwidth]{{./figs/efig1-Densities-EXP-OM7.5-467-468-582-593-612}.png}
%\captionsetup{labelformat=empty}
%\caption{\textbf{Fig. S1. Granulation of the lithium Bose-Einstein condensate, after $\mathbf{\tm=250}$ms and $\mathbf{\thd=250}$ms.} Five different experimental realizations (shots) corresponding to identical initial conditions, namely  $\Omega/2\pi=60Hz$ and $a(t)=8^{a_{+}=20}_{a_{-}=0.5} a_0$. 
% $B_{av}=714$G, $\delta{}B=41.3$G
%The fluctuations from shot to shot reveal that quantum correlations are large and underpin the emergence of granular structures.}
%\label{extfig1}
%\end{figure}

%\clearpage
%\begin{figure}
%\includegraphics[trim={0 0 0 0},clip, width=0.72\textwidth]{{./figs/efig2-Densities-EXP+GP-OM8.75-540-542-695-696-606}.png}
%\captionsetup{labelformat=empty}
%\caption{\textbf{Fig. S2. Long-time dynamics of grains.} Granulation is seen in the laboratory for a variety of modulating frequencies and for long times after the excitation has been switched off (also shown in Fig.~\ref{fig1} and Fig.~S1). Here $\Omega/2\pi=70Hz$ and $a(t)$ is the same as in Fig.~S1. The hold times are depicted whereas $\tm=250$ms in (a),(c) and $\tm=1000$ms in (b). Ecah image is a separate experimental realization.}
%\label{extfig2}
%\end{figure}
\renewcommand{\thefigure}{S\arabic{figure}} 
\setcounter{figure}{0} 
\newpage
\begin{figure}
%\vspace*{0mm}\includegraphics[width=0.72\textwidth]{{./figs/efig3a-Fit+Constrast-EXP}.png} \\~\\~
%\vspace{0mm}\includegraphics[width=0.72\textwidth,trim={0 0 0 28mm},clip]{{./figs/efig3b-Fit+Constrast-MB}.png}
\vspace*{0mm}\includegraphics[width=0.72\textwidth]{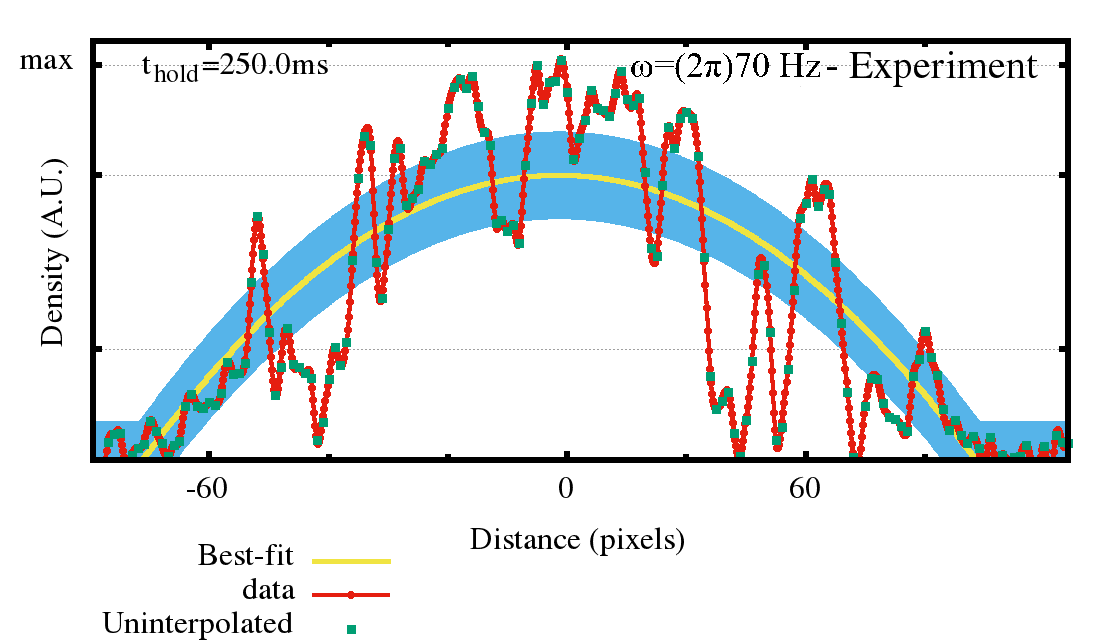} \\~\\~
\vspace{0mm}\includegraphics[width=0.72\textwidth,trim={0 0 0 28mm},clip]{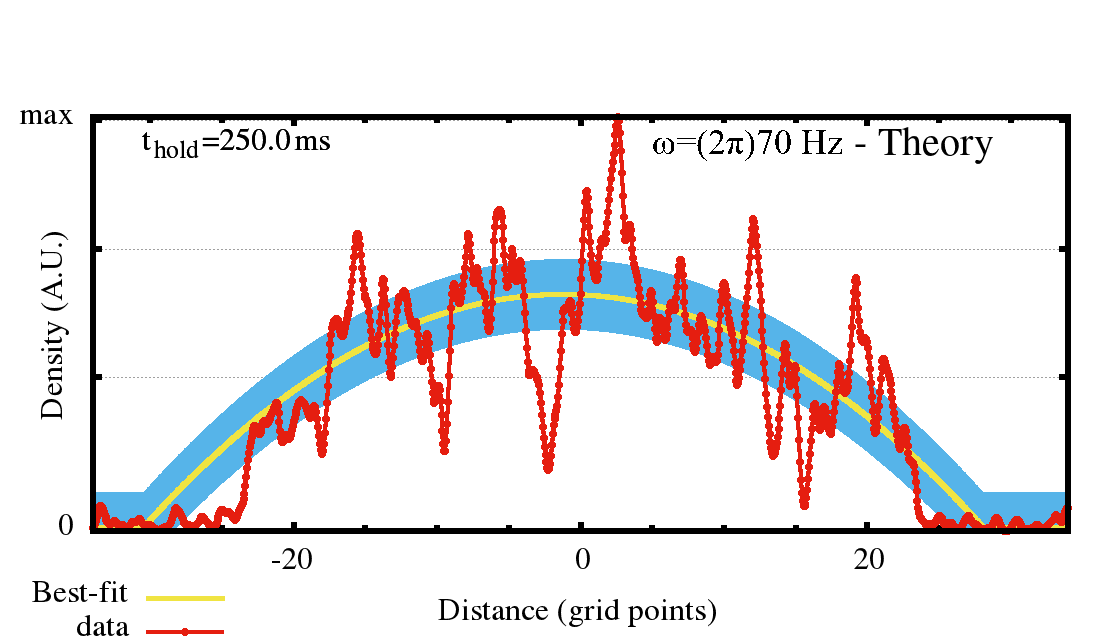}
\caption{Example of data fitting. (upper) Experimental and (lower) numerical data are fitted to a parabolic curve (yellow) in order to estimate $\mathcal D$ (see Methods). Only values of $\mathcal D$ that deviate more than $20\%$ from the value of the fitting function (i.e. points that lie outside the shaded area) are taken into consideration. The images are taken at $\Delta{\rm t}=\tm+\thd=250+250$ms. The numerical simulation is a $1$D model with $N=10^4$ and $M=2$ and the grid extension is [-128:128].}
\label{extfig3} 
\end{figure}

\newpage
\begin{figure}
\includegraphics[width=1\textwidth,angle=-90]{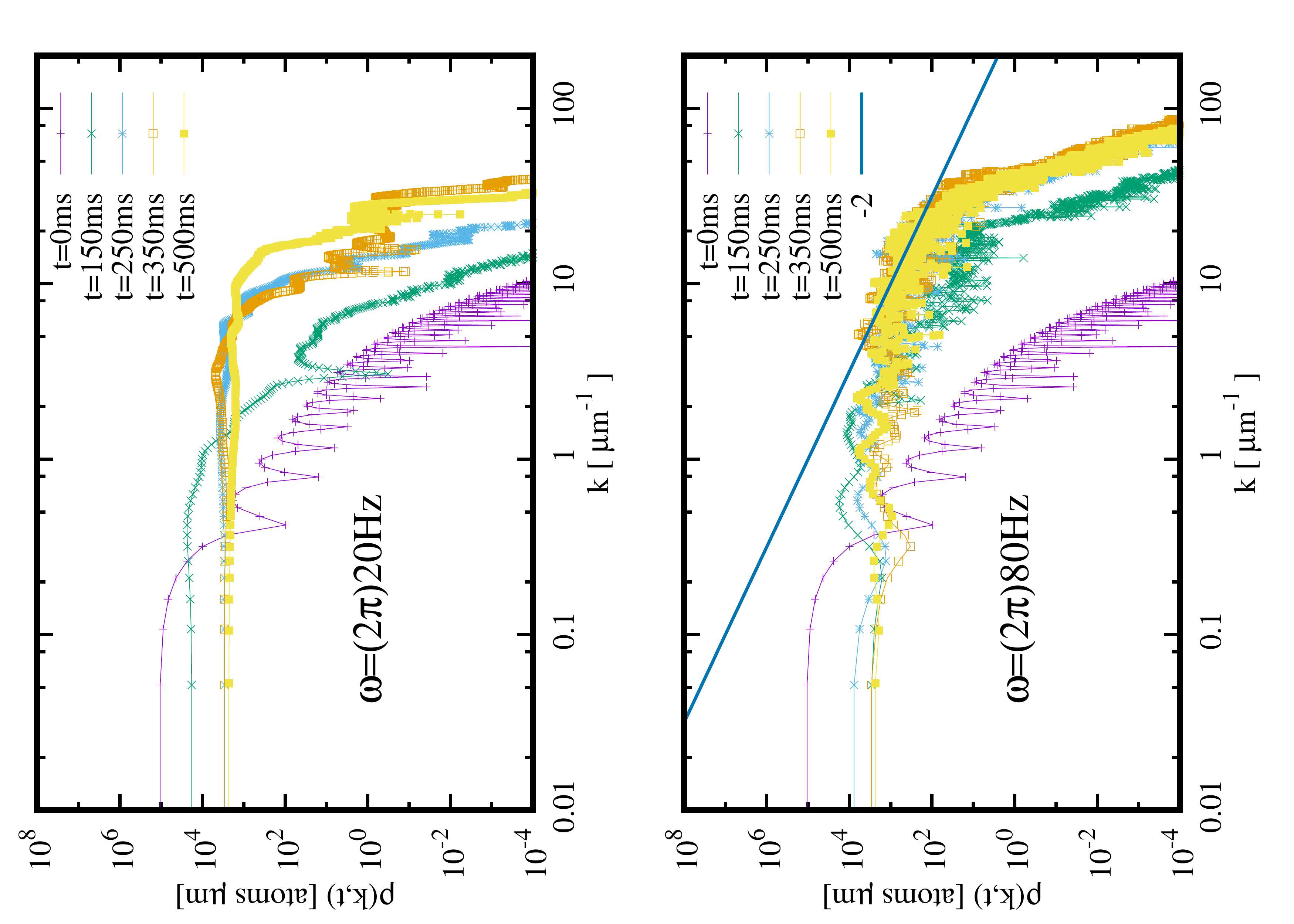}
%\captionsetup{labelformat=empty}
\caption{Density in momentum space. k-space densities for the regular (upper) and the granulated gas (lower panel) as calculated from the MB theory at different times (during modulation for $t\leq{}250$ms and after for $t>250$ms). In the granulated case the momentum distribution scales like $k^{-2}$ (straight line to guide the eye) for almost two decades, behavior that is characteristic of quantum turbulence. Contrary to the regular gas, this scaling remains even $250$ms after the modulation.} 
\label{extfig6}
\end{figure}

\newpage
\begin{figure}
\includegraphics[width=0.50\textwidth]{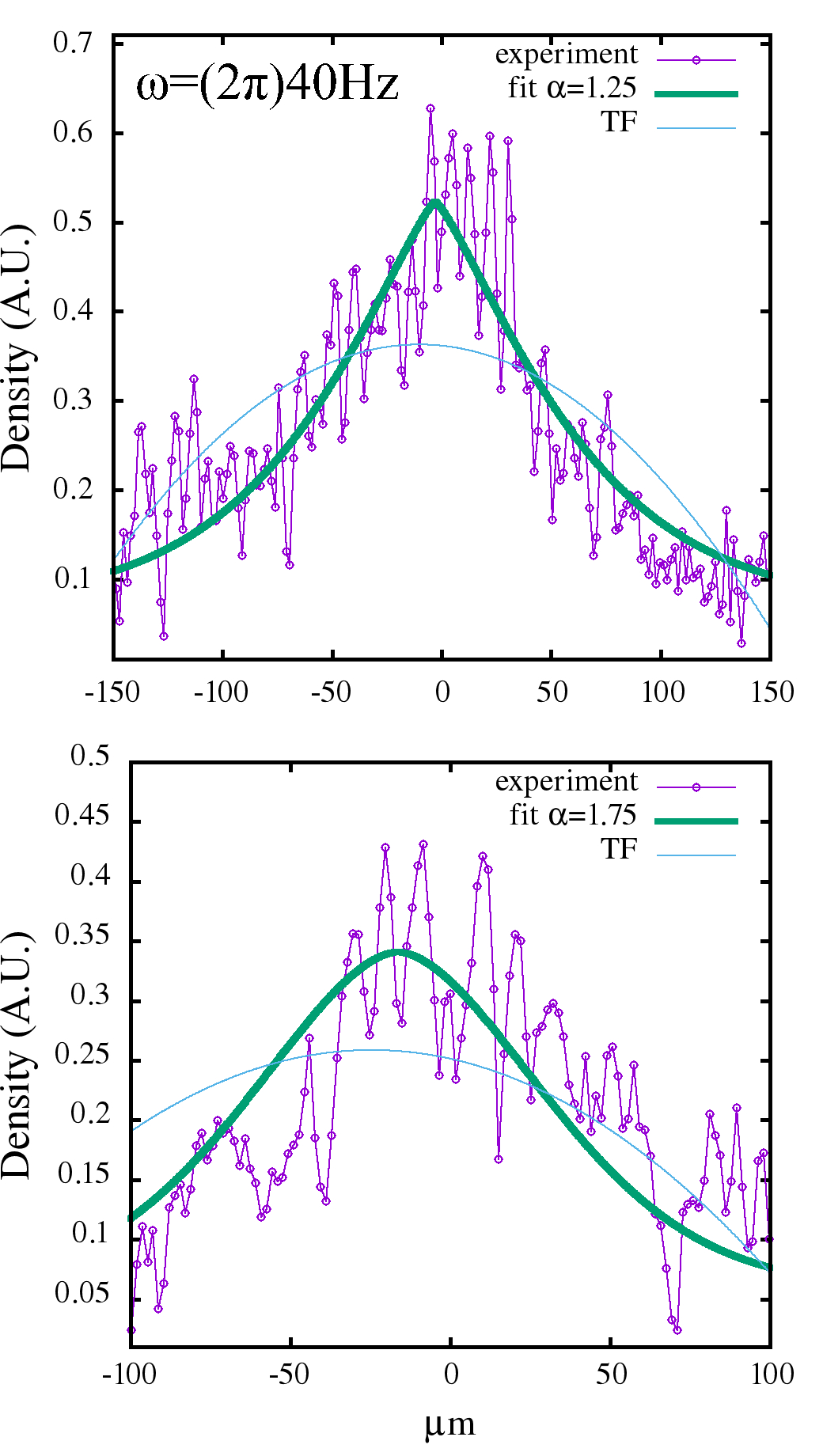}
%\captionsetup{labelformat=empty}
\caption{Experimental column densities exponentially fitted. Close to the threshold frequency $\omega/2\pi=40$Hz, where the system transitions from regular to granulated states, anomalous spatial distributions are seen (here, two experimental shots for the same initial conditions). These might bear resemblance to localized states, that have been shown to exist in BECs in optical lattices \cite{Roati2008}. We fit our observed density distributions to $C+A \exp\left(-\frac{|x-x0|^\alpha}{d}\right)$ and obtain $\alpha=1.25$ and $1.75$ for the two shots. The transition from a regular to a localized states happens as $\alpha\rightarrow 1$. For comparison, we plot the parabolic Thomas-Fermi (TF) fit (blue). 
}
\label{extfig7}
\end{figure}

\end{document}